\documentclass[12pt]{iopart}
\usepackage{iopams}
\usepackage[                
  retainorgcmds               
]{IEEEtrantools}
\usepackage{mathtools}
\usepackage{mathrsfs}
\usepackage{xparse} 
\usepackage{cite}
\usepackage{ulem}
\usepackage{braket}         
\usepackage{tikz}           
\usetikzlibrary{shapes.geometric, arrows}
\usepackage{circuitikz}
\newcommand{\matr}[1]{\ensuremath{\underline{\mathbf{#1}}}}
\newcommand{\uv}[1]{\ensuremath{\mathbf{\hat{#1}}}} 
\newcommand\omicron{o}
\newcommand\varpm{\mathbin{\vcenter{\hbox{%
  \oalign{\hfil$\scriptstyle+$\hfil\cr
          \noalign{\kern-.3ex}
          $\scriptstyle({-})$\cr}%
}}}}
\tikzstyle{startstop} = [rectangle, minimum width=3cm, minimum height=1cm,text centered, draw=black, fill=red!30]
\tikzstyle{io} = [rectangle, minimum width=3cm, minimum height=1cm, text centered, draw=black, fill=blue!30]
\tikzstyle{process} = [rectangle, minimum width=3cm, minimum height=1cm, text centered, draw=black, fill=orange!30]
\tikzstyle{decision} = [diamond, minimum width=3cm, minimum height=1cm, text centered, draw=black, fill=green!30]
\tikzstyle{arrow} = [thick,->,>=stealth]

\begin{document}
\title[Phonons by DFPT using the FLAPW method in \texttt{FLEUR}]{Phonons from Density-Functional Perturbation Theory using the All-Electron Full-Potential Linearized Augmented Plane-Wave Method \texttt{FLEUR}\footnote{Dedicated to the memory of Henry Krakauer (1947--2023)}}
\author{Christian-Roman Gerhorst$^{1,*}$, Alexander Neukirchen$^{1,2,*}$, Daniel A. Klüppelberg$^{1}$, Gustav Bihlmayer$^1$, Markus Betzinger$^1$, Gregor Michalicek$^1$, Daniel Wortmann$^1$ and Stefan Blügel$^1$
}
\address{$^1$ Peter Gr\"unberg Institute and Institute for Advanced Simulation, Forschungszentrum Jülich and JARA, 52425 Jülich, Germany}
\address{$^2$ Physics Department, RWTH-Aachen University, 52062 Aachen, Germany}
\begin{indented}
\item[] $^*$ Authors to whom any correspondence should be addressed.
\end{indented}
\eads{\mailto{c.gerhorst@fz-juelich.de}, \mailto{a.neukirchen@fz-juelich.de}, \mailto{g.bihlmayer@fz-juelich.de}}
\vspace{10pt}
\begin{indented}
\item[] \today
\end{indented}
\begin{abstract}
    Phonons are quantized vibrations of a crystal lattice that play a crucial role in understanding many properties of solids. Density functional theory~(DFT) provides a state-of-the-art computational approach to lattice vibrations from first-principles.
    We present a successful software implementation for calculating phonons in the harmonic approximation, employing density-functional perturbation theory (DFPT) within the framework of the full-potential linearized augmented plane-wave (FLAPW) method as implemented in the electronic structure package \texttt{FLEUR}.
    The implementation, which involves the Sternheimer equation for the linear response of the wave function, charge density, and potential with respect to infinitesimal atomic displacements, as well as the setup of the dynamical matrix, is presented and
    the specifics due to the muffin-tin sphere centered LAPW basis-set and the all-electron nature are discussed.
    As a test, we calculate the phonon dispersion of several solids including an insulator, a semiconductor as well as several metals. The latter are comprised of magnetic, simple, and transition metals. The results are validated on the basis of phonon dispersions calculated using the finite displacement approach in conjunction with the \texttt{FLEUR} code and the \texttt{phonopy} package, as well as by some experimental results.
    An excellent agreement is obtained.
\end{abstract}


\hfill

\submitto{IOP Electronic Structure}

\maketitle

\section{Introduction} \label{sec:intro}

Phonons are quantized collective lattice vibrations featuring a discrete spectrum of frequencies. They are also described and understood as quasiparticles using the framework of quantum field theory. The basic theory of phonons is well understood and has been described in detail in text books~\cite{BornHuang,Ashcroft76}. In the harmonic approximation, the phonon frequencies $\omega$ are determined by the eigenvalues 
    \begin{IEEEeqnarray}{rCl}
    \det \left|\matr{D}-\omega^2\matr{1}\right|\, =0
    \label{eq:DM-eigenvalue}
    \end{IEEEeqnarray}
    of the  dynamical matrix (DM) $\matr{D}$. $\matr{1}$ denotes the identity matrix. The dynamical matrix, here presented in terms of matrix elements  
    \begin{IEEEeqnarray}{rCl}
    D_{i'i}^{\gamma'\gamma}(\mathbf{q}) &=& \frac{1}{\sqrt{M_{\gamma'}M_{\gamma}}}\sum_{\mathbf{R}}\phi_{i'i}^{\gamma_{\mathbf{R}}'\gamma}e^{\mathrm{i}\mathbf{q}\cdot\mathbf{R}} ,
    \label{eq:DM}
    \end{IEEEeqnarray}
    with the atomic masses $M_{\gamma(')}$ is the reduced Fourier transform of the  harmonic force constant matrix $\matr{\phi}$, also known as Hesse matrix of second-order derivatives $\matr{E}^{(2)}_{\mathrm{tot}}$ with matrix elements
    \begin{IEEEeqnarray}{rCl}
    \phi_{i'i}^{\gamma'\gamma} &=& \frac{\partial^{2} E_{\mathrm{tot}}}{\partial\tau_{\gamma'i'}\partial\tau_{\gamma i}} = E^{(2)\gamma' i'\gamma i}_\mathrm{tot}
    \label{eq:FCM}
    \end{IEEEeqnarray}
    that describes consistent with the harmonic approximation the second-order  expansion of the Born--Oppenheimer energy $E_{\mathrm{tot}}$ with respect to the positions $\boldsymbol{\tau}$ of the atoms $\gamma$, and $\gamma^\prime$ in some unit cells. $\tau_i$ with $i\in\{1,2,3\}$ are the cartesian components of $\boldsymbol{\tau}$.  $\gamma_{\mathbf{R}}'$ describes the atom $\gamma'$ in the unit cell with lattice vector $\mathbf{R}$ at $\boldsymbol{\tau}_{\gamma_{\mathbf{R}}'}=\boldsymbol{\tau}_{\gamma'}+\mathbf{R}$. The symbol $\mathbf{q}$ denotes the phonon wave vector defined within the Brillouin zone (BZ) of the crystal lattice. It represents the momentum associated with a phonon and describes the propagation direction and the wavelength of the lattice wave. The dimension of the hermitian dynamical matrix scales with the dynamical degrees of freedom of the lattice, \textit{i.e.}\ the number of atoms in the unit cell, $N_\mathrm{A}$, along  the three cartesian coordinates as $\dim(\matr{D})=3N_{\mathrm{A}}\times 3N_{\mathrm{A}}$. The solutions of \eqref{eq:DM-eigenvalue}  are displacement modes 
     \begin{equation}
        \mathbf{w}^{\gamma_{\mathbf{R}}}_\mu(\mathbf{q}, t) = \frac{w^{\gamma_{\mathbf{R}}}_{\mu}(0)}{\sqrt{\mathrm{\Omega}_\mathrm{BZ}}}\frac{1}{\sqrt{M_\gamma}}\left[\mathbf{P}^{\gamma}_{\mu}(\mathbf{q})\mathrm{e}^{\mathrm{i}\left(\mathbf{q}\cdot \mathbf{R}+\omega_\mu(\mathbf{q}) t\right)} + \mathbf{P}^{*\gamma}_{\mu}(\mathbf{q})\mathrm{e}^{-\mathrm{i}\left(\mathbf{q}\cdot \mathbf{R}+\omega_\mu(\mathbf{q}) t\right)}\right] \,,
    \end{equation}
where the normalized polarization vector $\mathbf{P}^{\gamma}_{\mu}\in \mathbb{C}^3$ and $w^{\gamma_{\mathbf{R}}}_{\mu}(0)$ denotes  the direction and the arbitrary amplitude of the displacement for the $3N_\mathrm{A}$ phonon modes $\mu$ of atom $\gamma$ in unit cell $\mathbf{R}$ with wave vector $\mathbf{q}$ in Brillouin zone of volume $\mathrm{\Omega}_\mathrm{BZ}$.

    Phonons play a crucial role in understanding a vast number of material phenomena.
    They lie at the heart of thermodynamical properties of solids, heat and sound propagation, reveal elastic properties of materials, and contribute to electrical resistivity~\cite{BornHuang,Venkataraman,Srivastava,YuCardona}. In conventional superconductors, the interaction of electrons with phonons is the primary mechanism responsible for the attractive pairing of electrons leading to the superconducting state~\cite{PhysRev.106.162}.
    Phonons are of continuous interest
    due to their role in engineering acoustic metamaterials~\mbox{\cite{RHOlssonEl-Kady2008, LiNingLiuEtAl2020}}, as driving force for charge-density waves~\cite{doi:10.1073/pnas.1424791112}, for the optimization of the phonon transport in thermoelectrics~\cite{C5TC01670C}, and in the context of ferroelectric~\cite{Im2022}, 2D~\cite{graphene2D,Gu_2016,Gong_2022}, and magnetic~\cite{Gu_2022} materials. In the latter they contribute to spin-relaxation~\cite{Lunghi:22}, Gilbert damping~\cite{PhysRevB.99.184442} and -equilibration~\cite{MaehrleinRaduMaldonadoEtAl2018}, assist magnetization switching by linearly-~\cite{2021NatPh..17..489S} and circularly-polarized ~\cite{davies2023phononic}, or chiral phonons~\cite{Zhu2018-gk}, influence the temperature dependence of the magnetocrystalline anisotropy \cite{PhysRevLett.118.117201}, and can be of interest in the fields of orbitronics~\cite{GoJoLeeEtAl2021} and thermal Hall physics~\cite{PhysRevLett.123.167202}.
    
    Several theoretical approaches are employed to evaluate phonon properties in condensed matter systems~\cite{Chaplot}. Among those, the Kohn--Sham (KS) density functional theory (DFT)~\cite{Hohenberg_1964_PhysRev_136_864, KohnSham_1965_PhysRev_140_A1133, RevModPhys.71.1253, Becke2014, Jones2015} has established itself as the method of choice for providing materials specific information directly from the electronic structure without adjustable parameters. 
    Concerning the computation of phonons and related quantitites, there are basically two established DFT approaches in use: (i) the finite displacement (FD) method~\cite{KuncMartin1981, Kunc1982, PhysRevLett.69.2799}, and (ii) the density-functional perturbation theory (DFPT)~\cite{Zein1984, Baroni1987Green, Gonze1989, Baroni2001_RevModPhys_73_515, Gonze1997}.
    The FD method emerged first and has been preferred in a wide spectrum of the literature to this day~(see \textit{e.g.}\ \cite{oguchi}), while the number of publications using DFPT is constantly increasing. FD and DFPT are complementary to each other and, given the same input, are able to deliver quantitatively comparable results. They are often applied in parallel, \textit{e.g.}\ to test the degree of unharmonicity in teh FD results.  
    Beneficial for both methods is the $2n+1$-theorem~\cite{Gonze1989}, which gives access to quantities of order $2n+1$, while only having input quantities of order $n$ at hand. Both methods do not deliver continuous dispersion relations, which is why usually interpolation methods are performed as a post-processing step.
    However, the DFPT method excels in improving the interpolation at specific points of interest in the Brillouin zone, since it offers access to them at affordable numerical costs.
    A comprehensive overview is given in the review of 
    Baroni~\cite{Baroni2001_RevModPhys_73_515} or the book of 
    Martin~\cite{MartinESCambrUnivPress}.
    
    In the FD method, one takes advantage that the force acting on an atom due to a displacement is the first derivative of the total energy, $\mathbf{F}=-\partial E_\mathrm{tot}/\partial \boldsymbol{\tau}$, and the force-constant matrix elements
    \begin{IEEEeqnarray}{rCl}
        \phi_{i'i}^{\gamma'\gamma} = -\frac{\partial F_{\gamma i}}{\partial \tau_{\gamma'i'}}  \approx -\frac{{F_{\gamma i}}(\tau_{\gamma'i'} + \mathrm{\Delta}{\tau_{\gamma' i'}}) - {F_{\gamma i}}(\tau_{\gamma'i'})}{\mathrm{\Delta}{\tau_{\gamma'i'}}}
    \end{IEEEeqnarray}
    are calculated by a difference quotient of the $i$-th Cartesian component of the force $F_{\gamma i}$ acting on atom $\gamma$ when another atom $\gamma'$ of the solid is displaced by a small displacement $\mathrm{\Delta}\tau_{\gamma'i'}$ from the equilibrium position $\boldsymbol{\tau}_{\gamma'}$ into direction $i'$. At equilibrium $F_{\gamma i}(\tau_{\gamma'i'})$ is usually zero.
    For crystalline solids, on which we focus throughout the paper, symmetry can usually be exploited, reducing the number of necessary force-vector components and displacements. Additionally, the combinations of $(\gamma',\gamma)$ reduce to the $N_{\mathrm{A}}^{2}$ pairs of atoms $(\beta,\alpha)$ in the representative unit cell.
    This is normally automated by software packages such as \texttt{phonopy}~\cite{phonopy-phono3py-JPCM,phonopy-phono3py-JPSJ}, which provide  phonon calculations at harmonic and quasi-harmonic levels. 
     Overall, the implementation of the FD method is quite simple, provided the DFT code delivers reliable forces.
    Nevertheless, the supercells must be chosen to include different periods determined  by the phonon vector~$\mathbf{q}$.
    As a consequence, the~$\mathbf{q}$-vector must be commensurate to the supercell, restricting this method to rational~$\mathbf{q}$-vector components, and making the calculation of phonons with a $\mathbf{q}$-vector exhibiting a small absolute value very expensive.
    
    In DFPT, we take an analytical approach to the second derivates of the total energy. Then, the DM contains, among other terms, linear responses of the charge density and the effective potential to the change of the external potential caused by the phonon. In DFPT, the first-order response functions of the electronic structure to small perturbations of the atom positions without the need to perform completely new calculations for each perturbation are calculated in a self-consistent way using the Sternheimer equation~\cite{Sternheimer1954}, which is a first-order version of the KS eigenvalue equation. As will be outlined in subsection~\ref{sec:dfpt}, this gives access to $E_{\mathrm{tot}}^{(2)}$ avoiding supercell calculations. The costs of a DFPT calculation are equally distributed among arbitrary $\mathbf{q}$-vectors and comparable to a DFT self-consistency procedure.

Most phonon studies using DFPT have been performed with norm-conserving pseudopotentials~\cite{GonzeBeukenCaracasEtAl2002,Segall2002,GiannozziBaroniBoniniEtAl2009,Andrade2012}, but there are now also a plethora of studies using  ultrasoft (US) pseudopotentials~\cite{PhysRevB.64.235118,PhysRevB.71.115106,PhysRevB.76.054308,PhysRevB.100.045115} and the projector-augmented wave (PAW) method~\cite{PhysRevB.81.075123,PhysRevB.82.075116}. Publications and codes combining DFPT and all-electron muffin-tin based electronic structure methods such as the augmented spherical wave (ASW)~\cite{eyert2013planewave}, linear muffin-tin orbital techniques (LMTO)~\cite{Questaalcode2020}, Korringa--Kohn--Rostoker (KKR) Green function~\cite{Papanikolaou2002}, and full-potential linearized augmented plane-wave (FLAPW) method~\mbox{\cite{Weinert1981_JMathPhys_22.2433, WimmerKrakauerWeinertEtAl1981, WeinertWimmerFreeman1982, NICFLAPWBlügelBihlmayer}} are scarce~\cite{Savrasov1996_PhysRevB_54_16470, PhysRevB.49.4467,Kouba2001} and the technicalities of the implementation are not well explored. All-electron methods treat core and valence electrons on the same footing.  
In  order to deal with the Coulomb singularity produced by the nuclear charge and the associated rapid variation of the core and valence electron wave functions and charge densities in the vicinity of the nucleus, all-electron methods partition the space of the unit cell into muffin-tin spheres in which wave functions, charge densities and potentials are represented in real space. A Fourier representation of these quantities would hardly converge. 

In this paper, we present a successfully working implementation of DFPT in the context of the all-electron FLAPW method. The FLAPW method is frequently considered a reference for electronic structure (DFT) calculations~\cite{doi:10.1126/science.aad3000, bosoni2023verify}, especially when dealing with magnetism, systems with localized electrons such 2p, 3d, and 4f electrons, or open systems, and systems  in lower dimensions. The FLAPW methodology is well-developed~\cite{singhnordstrom} and first-order changes of the total energy such as forces~\cite{Soler:1989, Soler:1990, Yu1991_PhysRevB_43_6411} or the stress-tensor~\cite{Belbase:21} are well-established. The second-order changes, however, are at a different scale, since they also require the density response in the form of first-order changes of the density and second-order derivatives, which require greater numerical attention as differentiation acts numerically as a roughening operator. Here, we present solutions to known numerical challenges of muffin-tin based electronic structure methods in general, and the FLAPW method in particular, in the context of the DFPT approach such as:
The Madelung summation, the Coulomb singularity of the potential, the rapidly varying wave functions and charge densities in the vicinity of the nucleus, the calculations of gradients of the all-electron potential, the presence of the core electrons, the incompleteness and the position dependence of the basis-set, the different representations of the basis-set in muffin-tin-spheres and the interstitial region and their match at the muffin-sphere boundary. On a more general level, this implementation allows to gain insight in response properties of highly complex materials. 

We implemented our approach in the open source electronic structure package \texttt{FLEUR}~\cite{fleurWeb, fleurCode}, more precisely in the bulk version of general symmetry. In the context of this work, it is worth mentioning that an emphasis was placed on the implementation of a numerically accurate force formalism~\cite{Yu1991_PhysRevB_43_6411, PhysRevB.91.035105} (to which the DFPT implementation is very alike to), and on the choice of local orbitals~\cite{SJOSTEDT200015} to reduce the linearization error~\cite{Friedrich:06,MICHALICEK20132670} and to improve the LAPW basis set~\cite{D_D_Koelling_1975} towards unoccupied states of higher energies~\cite{PhysRevB.83.045105}.
We show that the  implementation of the DFPT presented here, which is based on the dissertations of Klüppelberg~\cite{KlueppelbergDr} and Gerhorst~\cite{GerhorstDr:}, in which further nitty-gritty details can be found, 
provides a solid foundation for calculating the phononic properties, and charge density response properties in general of complex materials with the FLAPW method according to first principles.
    
    This paper is organized as follows: We briefly recapitulate the central theoretical background of DFT, DFPT, and the FLAPW method, also to establish a consistent notation.
    We then explain the technical details of our implementation. We present the general concept and the workflow for DFPT calculations, discuss the challenges related to the choice of the LAPW basis, the implementation of the Sternheimer equation and the dynamical matrix, and solutions to the challenges. In order to guarantee a good reading flow of the paper and not to be overloaded with details, we have separated additional technical details into  \ref{sec:app_psi1} to \ref{sec:app_occ}. Although for clarity and simplicity the implementation is presented based on electronic charge density only as in the context of non-spin-polarized DFT, we also apply this method to collinear magnets by doubling the formalism and incorporating the magnetization density, thus replacing the charge density by spin-densities of spin-up and -down electrons.  
    Finally, we validate our DFPT framework with respect to the quality of Goldstone modes and phonon dispersion relations against the FD approach on a selection of materials and conclude with  an outlook to future developments.
\section{Theoretical Background} \label{sec:theory}
    \subsection{Density Functional Theory}
    \label{ssec:DFT}
    According to the Kohn--Sham DFT, the  total energy of a system of interacting electrons is uniquely determined by its ground-state charge-density distribution and the problem of a system of interacting electrons is  mapped onto an equivalent non-interacting problem with same ground-state density. This is made possible by expressing the unknown density functional in the form 
    \begin{IEEEeqnarray}{rCl}
    E[n]= T_0[n]+E_\mathrm{H}[n]+E_\mathrm{xc}[n] +\int V_\mathrm{ext}\left(\{\boldsymbol{\tau}\},\mathbf{r}\right)n(\mathbf{r})\mathrm{d}\mathbf{r} + E_{\mathrm{ion-ion}}(\{\boldsymbol{\tau}\})\,,
    \label{eq:KS-energy}
    \end{IEEEeqnarray}
    where the first term is the kinetic energy of the non-interacting system, the  second term is the classical electrostatic self-interaction of the electron charge-density distribution, known as Hartree energy $E_\mathrm{H}$, the third  is the unknown and well-approximated exchange-correlation (xc) energy $E_\mathrm{xc}$, and the fourth term  describes the interaction of electrons with the potential external to the electrons, \textit{e.g.}\ of the nuclei positioned at  $\boldsymbol{\tau}$. The final term describes the electrostatic interaction among the nuclei.
    This approach is in principle exact, but the aforementioned exchange-correlation energy is not known explicitly and there exists a large variety of approximations~\cite{Rappoport2009,burkereview}.
    In this paper we work with the local-density approximation~\cite{vwn}, a simple representative of the xc-functionals, which leads to good results for a large class of materials.
    For a set of atoms located at  $\{\boldsymbol{\tau}\}$, the ground-state density is obtained through the Kohn--Sham equations 
    \begin{IEEEeqnarray}{rCl}
        \left(- \frac{\Delta}{2} + {V_{\mathrm{eff}}}(\{\boldsymbol{\tau}\},\mathbf{r})\right) {\Psi_{\nu}}(\mathbf{r}) = \epsilon_{\nu} {\Psi_{\nu}}(\mathbf{r}) \IEEEyesnumber \IEEEyessubnumber \label{eqn:KSeqA}
    \end{IEEEeqnarray}
    \vspace{-.5cm}
    \begin{IEEEeqnarray}{rCl}
      {V_{\mathrm{eff}}}[{n^{(0)}}](\{\boldsymbol{\tau}\},\mathbf{r}) \coloneq {V_{\mathrm{ext}}}(\{\boldsymbol{\tau}\},\mathbf{r}) + V_{\mathrm{H}}[{n^{(0)}}](\mathbf{r}) + {V_{\mathrm{xc}}}[{n^{(0)}}](\mathbf{r}) \IEEEeqnarraynumspace \IEEEyessubnumber
    \end{IEEEeqnarray}
    \vspace{-.5cm}
    \begin{IEEEeqnarray}{rCl}
       {n^{(0)}}(\mathbf{r}) = \sum_{\omicron} |{\Psi_{\omicron}}(\mathbf{r})|^{2} \IEEEyessubnumber \label{eqn:kohn_sham_electronicDensity} 
    \end{IEEEeqnarray}
    that are solved self-consistently and comprise the effective potential $V_{\mathrm{eff}}$, being a functional of the ground-state charge density $n^{(0)}(\mathbf{r})$, subdivided into a sum of the external (ext), Hartree (H), and exchange-correlation potential (xc). The external and Hartree apart are often grouped as the Coulomb potential $V_{\mathrm{C}}$. ${\Psi_{\nu}}$ are the eigenstates and $\epsilon_{\nu}$ the eigenenergies.
    The index $\nu$ ($\omicron$) denotes (occupied) spin-degenerate states. 
    
    For simplicity, throughout this work the spin index is omitted. For magnetic systems, we switch to the well-established spin-density functional theory~\cite{UvonBarth1972}, where the treatment of collinear magnets is straightforward: The spin-degeneracy is lifted and~\eqref{eqn:KSeqA} is solved separately for the spin-up ($\uparrow$) and -down ($\downarrow$) states, $\Psi_{\nu\uparrow(\downarrow)}$,  solutions of a spin-dependent potential $V_{\mathrm{eff},\uparrow(\downarrow)}=V_{\mathrm{ext}}+V_{\mathrm{H}}[n^{(0)}]+V_{\mathrm{xc},\uparrow(\downarrow)}[n_{\uparrow}^{(0)},n_{\downarrow}^{(0)}]$.
    The latter term is obtained by generalizing the ground-state  density to the ground-state  spin densities $n_{\uparrow (\downarrow)}^{(0)}$, calculated via the summations~\eqref{eqn:kohn_sham_electronicDensity} of spin-up and -down states, separately, with $n^{(0)}= n_{\uparrow}^{(0)}+n_{\downarrow}^{(0)}$. The spin-dependent exchange correlation potential $V_{\mathrm{xc},\uparrow(\downarrow)}$ is related to the spin-independent exchange correlation potential $V_{\mathrm{xc}}$ and a magnetic exchange-correlation field $B_{\mathrm{xc}}$  as  $V_{\mathrm{xc},\uparrow(\downarrow)}[n_{\uparrow}^{(0)},n_{\downarrow}^{(0)}] =V_{\mathrm{xc}}[n_{\uparrow}^{(0)},n_{\downarrow}^{(0)}]\varpm B_{\mathrm{xc}}[n_{\uparrow}^{(0)},n_{\downarrow}^{(0)}]$. These generalizations hold also true for the phonon calculations below.
    \subsection{Density-Functional Perturbation Theory} \label{sec:dfpt}
        Depending on the energy scales or phenomena of interest, quantities of certain orders in a perturbation are to be determined.
        Given a phonon, the dynamical matrix and consequently the second-order changes in the total energy, $\matr{E}_{\mathrm{tot}}^{(2)}$, with respect to atomic displacements $\partial \boldsymbol{\tau}$ turn out to be pivotal. Applying the Hellmann--Feynman Theorem to the second order derivative of the energy $E$ in \eqref{eq:KS-energy} and restricting ourselves here for simplicity to one displacement component of one atom $\lambda =\tau_{\gamma i}$
        the second-order change of the energy reads
        \begin{IEEEeqnarray}{rCl}
            E_{\mathrm{tot}}^{(2)} =\frac{\mathrm{d}^2 }{\mathrm{d}\lambda^2}E_\mathrm{tot}&=& \int_{\mathrm{\Omega}}  \left({n^{(1)}}(\mathbf{r}) {V_\mathrm{ext}^{(1)}}(\mathbf{r}) + {n}(\mathbf{r}) {V_\mathrm{ext}^{(2)}}(\mathbf{r})\right) \mathrm{d}\mathbf{r} + E_{\mathrm{ion-ion}}^{(2)} \,, \label{eqn:verysimpleDM}
        \end{IEEEeqnarray}
        where the basis-set independent variation of the ion-ion interaction $E_{\mathrm{ion-ion}}^{(2)}$ is included. The integral spans over the volume of the unit cell $\mathrm{\Omega}$. In our nomenclature, quantities with the superscript $(1)$ (or $(2)$) are defined as perturbed quantities to first (second) order, while ones without a superscript are the unperturbed quantities of the ground-state system. Unlike first-order changes of the energy, such as forces or stress tensors, for second-order changes, the terms involving the derivative of the density do not vanish. This means that it is necessary to compute the electronic response of the system to the displacement of atoms to perform \textit{ab initio} lattice dynamics calculations. 
        The requirement of a first-order density change makes the calculation of quantities requiring second-order energy derivatives qualitatively very different from the evaluation of quantities requiring only first-order energy changes. 
        The first-order change in the density, $n^{(1)}$,  constitutes a key quantity of the DFPT and reads
       \begin{IEEEeqnarray}{rCl}
            {n^{(1)}_{\vphantom{{p}}}}(\mathbf{r}) &=& \sum_{\omicron}  {\Psi_{\vphantom{f}\omicron}^{*(1)}}(\mathbf{r}){\Psi_{\vphantom{f}\omicron}}(\mathbf{r})+ {\Psi_{\vphantom{f}\omicron}^{*}}(\mathbf{r}) {\Psi_{\vphantom{f}\omicron}^{(1)}}(\mathbf{r}) \label{eqn:rho1BasicTRS}
            \quad\mathrm{with}\quad N^{(1)}=\int_{\mathrm{\Omega}} n^{(1)}_{\vphantom{{p}}}(\mathbf{r})\,\mathrm{d}\mathbf{r}=0  \,,
       \end{IEEEeqnarray}
       relating to the first-order change in the eigenfunctions $\Psi_{\vphantom{f}\omicron}^{(1)}$. For our current implementation, exploiting symmetry this can be simplified further (see~\ref{sec:app_psi1}). The relationship between the second order change in the energy and the first-order change in the eigenvalues and eigenstates of the underlying Hamiltonian satisfies the well-known $2n+1$-theorem~\cite{Gonze1989}, which states that a $(2n+1)$-order derivative of the energy of some Hamiltonian can be calculated from the knowledge of the eigenfunction and its derivatives up to order $n$.
       
       Access to the aforementioned first-order change of the electronic quantities is provided by the solution of the Sternheimer equation in a self-consistent fashion, as the change in the charge density creates a change in the effective potential and vice versa. 
       Assuming non-degenerate states, the following basic form of the Sternheimer equation holds
        \begin{IEEEeqnarray}{rCl}
            \left(\mathscr{H}_{\vphantom{f}} - \epsilon_{\vphantom{f}o}\right) |\Psi_{\vphantom{f}o}^{(1)}\rangle = - \sum_{u} |\Psi_{\vphantom{f}u}\rangle \langle\Psi_{\vphantom{f}u}|V^{(1)}_{\mathrm{eff}}|\Psi_{\vphantom{f}o}\rangle \, , \IEEEeqnarraynumspace \label{eqn:verysimpleSH} 
        \end{IEEEeqnarray}
        where $\mathscr{H}$ is the Hamiltonian and $V^{(1)}_{\mathrm{eff}}$ is the first-order change of the effective potential, which contains not only $V^{(1)}_{\mathrm{ext}}$ but also the Hartree and exchange correlation kernel, $\delta(V_{\mathrm{H}} + V_{\mathrm{xc}})/\delta n$, and the density response $n^{(1)}_{\vphantom{{p}}}$.  
        The projector onto the unoccupied subspace of states denoted by subscript $u$ is explicitly included. This gives rise to a self-consistency calculation: The potential response determines the response of the eigenstates, which are used to calculate the density response, that in turn is used to construct the potential response. This Sternheimer self-consistency cycle is very similar to that of a DFT ground-state calculation. The generation of the density and potential is replaced by the generation of their respective responses, the starting perturbation is only that of the external potential, instead of the original  Hamiltonian and overlap matrices the corresponding response matrices are set up and the density response is mixed to achieve self-consistency instead of the density itself. A key difference is that there is no diagonalization step as for the Schrödinger equation (solving the Sternheimer equation is purely matrix-vector multiplication), and that we need access to the full eigenspectrum of each k-point, not only the occupied states. After self-consistency is reached, the variational solution can be used to calculate the density response~\eqref{eqn:rho1BasicTRS} and subsequently the force constant matrix \eqref{eqn:verysimpleDM}.
        The ramifications of applying the formalism in the LAPW basis~\cite{D_D_Koelling_1975} will be explored in section~\ref{sec:implementation}. 
        
        In summary, the DFPT for phonons requires the first-order changes in the density, of the wave function, the external and effective potential, as well as the second order changes in the external potential (whose evaluation will be avoided in practical implementation) and the ion-ion energy. In reality, all first-order changes are vector quantities and all second-order changes are matrices, and the product $n^{(1)} V_\mathrm{ext}^{(1)}$ in \eqref{eqn:verysimpleDM} turns into a direct product $\mathbf{n}^{(1)}\otimes \mathbf{V}_\mathrm{ext}^{~(1)}$. Analogously to the discussions in \ref{ssec:DFT}, for collinear magnetic systems the Sternheimer equation~\eqref{eqn:verysimpleSH} is solved for the changes of the spin-up and -down states used to synthesize the spin-density response~\eqref{eqn:rho1BasicTRS}, which sum to the required density response. In principle,~\eqref{eqn:verysimpleDM} and~\eqref{eqn:verysimpleSH} are sufficient in a plane-wave ansatz and together with~\eqref{eqn:rho1BasicTRS} they make up the concept of this DFPT implementation. 
        However, the position dependent, incomplete and multi-domain represented basis-set of the LAPW basis gives rise to a multitude of additional terms, each of them to be carefully taken into account.
    \subsection{Full-Potential Linearized Augmented Plane-Wave Method} \label{sec:flapw}
        When describing wave functions, $\Psi_{\mathbf{k}\nu}(\mathbf{r})$, in a periodic lattice, their natural form are Bloch waves characterized by a crystal momentum vector $\mathbf{k}$ restricted to the first Brillouin zone (BZ) of the reciprocal lattice, and a band index $\nu$. They are typically expanded into basis functions, \textit{e.g.}\ plane waves, 
        with reciprocal lattice vectors $\mathbf{G}$, 
        \begin{IEEEeqnarray}{rCl}
            \Psi_{\mathbf{k}\nu}(\mathbf{r}) = \sum_{\mathbf{G}} z_{\mathbf{k}+\mathbf{G},\nu} \phi_{\mathbf{k}+\mathbf{G}}(\mathbf{r}) \qquad\forall\mathbf{k}\in\mathrm{BZ}\quad\mathrm{and}\quad |\mathbf{k}+\mathbf{G}|\le K_{\mathrm{max}} \, ,
        \end{IEEEeqnarray}
        where $z_{\mathbf{k}+\mathbf{G},\nu}$ are the corresponding expansion coefficients. The maximum length of the reciprocal lattice vectors, $K_{\mathrm{max}}$, determines the number of basis functions $N_\mathrm{B}$ and controls the numerical effort and precision of the results. 
        Care has to be taken when selecting the k-point set to maintain the symmetry of the lattice: here, we always choose an equidistant mesh containing the $\mathrm{\Gamma}$-point with $N_{k_x}\times N_{k_y}\times N_{k_z}$ k-points in the reciprocal space. For odd $N_{k_{i}}$ this corresponds to a Monkhorst-Pack mesh~\cite{PhysRevB.13.5188}. For further details on the choice of the k-point mesh see also section~\ref{ssec:CD-DFPT}.
        
        To deal with the Coulomb singularity $\sim\! 1 / r$ at the center of the atoms due to the positively charged nuclei and the rapidly oscillating core and valence electron wave functions in the vicinity of the nuclei, as typical for all-electron methods, in the FLAPW method~\cite{PhysRevB.24.864}, the computational domain is divided into spheres $\mathrm{MT}^{\gamma}$ around the centers of each atom $\gamma$ --- the union of all these spheres is called the muffin-tin (MT) region --- and into an interstitial (IR) region. 
        The basic plane-wave approach is kept in the interstitial region of the unit-cell with volume $\mathrm{\Omega}$, but it is augmented by radial functions $u_\ell(r)$ and spherical harmonics $Y_L(\mathrm{\hat{r}})$ with the angular momentum and magnetic quantum numbers $L=(\ell,m)$, and the unit vector $\mathrm{\hat{r}}=\mathbf{r}/r$ in the MT region.
        To guarantee sufficient variational flexibility of these LAPW basis functions, different "orders" (denoted by the index $p$) of radial functions are used.   
        The zeroth order $u_{\ell 0}(r)$ corresponds to the solution of the radial Schrödinger equation for a spherical potential (containing the full atomic $\sim 1 / r$ singularity) in the MT spheres to a given energy parameter characteristic of the valence electrons, and the first order functions $u_{\ell 1}(r)$ correspond to their first order energy derivatives~\cite{D_D_Koelling_1975}.
        Extending this logic, in certain cases, we supplement the LAPW basis ~\cite{D_D_Koelling_1975} with local orbitals~\cite{singh, PhysRevB.83.045105}.
        These are used to give more variational freedom, to accurately describe semicore states, high-lying unoccupied states, as well as to reduce the linearization error~\cite{MICHALICEK20132670}. These are exclusively present in the MT.
        The LAPW basis functions are thus
        \begin{IEEEeqnarray}{rCl}
            \phi_{\mathbf{k}+\mathbf{G}}(\mathbf{r}) =
            \begin{cases}
                \frac{1}{\sqrt{\mathrm{\Omega}}} \exp(\mathrm{i} (\mathbf{k} + \mathbf{G}) \cdot \mathbf{r}) , & \mathbf{r} \in \mathrm{IR} \\[10pt]
                \displaystyle{\sum_{Lp}} a^{\mathbf{k}+\mathbf{G},\gamma}_{Lp} u^{\gamma}_{\ell p}(r_{\gamma}) Y_{L}(\mathrm{\hat{r}}_\gamma), & \mathbf{r} \in \mathrm{MT}^{\gamma}
            \end{cases} \,, 
            \label{eqn:nVexpansion}
        \end{IEEEeqnarray}
with the unit cell volume $\mathrm{\Omega}$ and coefficients $a^{\mathbf{k}+\mathbf{G},\gamma}_{Lp}$, with $\ell\le \ell_\mathrm{max}\simeq R_{\mathrm{MT}^\gamma}K_\mathrm{max}$, 
determined such as to guarantee continuity and smoothness of the basis function at the muffin boundary. $\ell_\mathrm{max}$ is a numerical cut-off parameter often set by the muffin-tin radius $R_{\mathrm{MT}^\gamma}$ and the largest reciprocal lattice vector controlling the quality of the basis, $K_\mathrm{max}$. The radial functions are represented at a set of $N_{\mathrm{MT}}$ radial mesh points. 
        
        Consequently, it is natural to choose the computational domain consisting of MT and IR also for the charge density, ${n}(\mathbf{r})$, and potential, $V_{\mathrm{eff}}(\mathbf{r})$, and expand both into plane waves (up to a maximal wave vector length $G_{\mathrm{max}}$) and radial functions times spherical harmonics (up to an angular quantum number $L_{\mathrm{max}}\le 2\ell_\mathrm{max}$), as exemplified here for the densities:
        \begin{IEEEeqnarray}{rCl}
            {n}(\mathbf{r})=
            \begin{cases}
                \sum_{\mathbf{G}} {n^{\mathrm{IR}}}(\mathbf{G}) \mathrm{e}^{\mathrm{i} \mathbf{G} \cdot \mathbf{r}}, & \mathbf{r} \in \mathrm{IR}
                \\[10pt]
                \sum_{L} n^\gamma_{L}(r_{\gamma}) Y_{L}(\mathrm{\hat{r}}_\gamma) & \mathbf{r} \in \mathrm{MT}^{\gamma} \,.\label{eq:simplyrho}
            \end{cases}
        \end{IEEEeqnarray}
     In practice, the support of the different regions is mediated by step functions $\mathrm{\Theta}_{\gamma}$ and $\mathrm{\Theta}_{\mathrm{IR}}$. $\mathrm{\Theta}_{\gamma}$ are $1$ in the respective $\mathrm{MT}^\gamma$ sphere of atom $\gamma$ and $0$ everywhere else. The step function $\mathrm{\Theta}_{\mathrm{IR}} = 1 - \sum_{\gamma} \mathrm{\Theta}_{\gamma}$ removes the MT region from its integration. The Fourier representation $\mathrm{\Theta}_{\gamma}(\mathbf{G})$ can be found in equation 5.41 of reference~\cite{singhnordstrom}.
        As the representation of the electronic structure in the MT spheres now explicitly depends on the atomic positions, several amendments to the previously outlined theory of phononic perturbations become necessary.
        These will be discussed in the next chapter. For practical purposes, the symmetry properties of the crystal lattice are used and in  the \texttt{FLEUR} code~\cite{fleurWeb, fleurCode}, the charge density and potential is represented in terms of symmetrized plane waves, so-called star-functions~\cite{singhnordstrom}, and symmetrized spherical harmonics, so-called  lattice harmonics~\cite{Altmann:65}. For the sake of readability, we largely omit this additional layer of representation in this work.
        
        In the LAPW basis, the Kohn--Sham equation \eqref{eqn:KSeqA} turns into a generalized eigenvalue problem 
         \begin{IEEEeqnarray}{rCl}
         \left(\matr{H}_{\mathbf{k}} - \epsilon_{\mathbf{k}} \matr{S}_{\mathbf{k}}\right)\mathbf{z}_{\mathbf{k}}= 0 \qquad\forall\mathbf{k}\in\mathrm{BZ}
         \label{eqn:GenEigenvalueproblem}
         \end{IEEEeqnarray}
         with a Hermitian Hamiltonian and overlap matrix of dimension $N_\mathrm{B}\times N_\mathrm{B}$ of matrix elements $H^{\mathbf{G}^\prime \mathbf{G}}_{\mathbf{k}}=\Braket{\phi_{\mathbf{k} + \mathbf{G}' } | \mathscr{H} | \phi_{\mathbf{k}+\mathbf{G}}}_{\!\!\mathrm{\Omega}}$ and $S^{\mathbf{G}^\prime \mathbf{G}}_{\mathbf{k}}=\Braket{\phi_{\mathbf{k} + \mathbf{G}' } | \phi_{\mathbf{k}+\mathbf{G}}}_{\!\!\mathrm{\Omega}}$, respectively. The setup of the Hamiltonian matrix due to the non-spherical potential $V_{L^\prime}$ in each MT sphere expressed as in \eqref{eqn:nVexpansion} comprises in the order of  $\mathcal{O}(N_\mathrm{A}(\ell_\mathrm{max}+1)^4)$ matrix elements of the type $\langle L|V_{L^\prime}|L^{''}\rangle$, \textit{i.e.}\ between the basis functions in spherical representation with the angular and magnetic quantum number $L$ and  $L^{''}$ and the non-spherical  potential. In order to exclude matrix elements of irrelevant magnitude we introduce an angular momentum cut-off $\ell_{\mathrm{max,nsph}}\le \ell_\mathrm{max}$ for the basis functions $\ell$ and $\ell^{''}$ contributing to the Hamiltonian setup.
         The diagonalization of the eigenvalue problem \eqref{eqn:GenEigenvalueproblem} is the runtime determining step of a self-consistent determination of the ground-state charge density, making the runtime of LAPW methods scale $\propto N_{\mathbf{k}}\times N_\mathrm{B}^3$, where scaling with respect to the number of basis functions, $N_\mathrm{B}$, stands also for the precision scaling of the physical properties as well as the volume scaling $\mathcal{O}(N_\mathrm{A}^3)$, as the number of basis functions scale linearly with the number of atoms, $N_\mathrm{B}\propto N_\mathrm{A}$.
\section{Implementation} \label{sec:implementation}
     The central motivation is to determine the phonon dispersion~\eqref{eq:DM-eigenvalue} in the harmonic approximation by means of the force constant matrix~\eqref{eqn:verysimpleDM}, which requires the determination of the charge-density response, the wave function response, the response of the external and effective potential, as well as the second order changes in the external potential and the ion-ion energy. In the following outline of the implementation we deal with vectors and matrices in the space of the $3N_\mathrm{A}$ dynamical degrees of freedom and of the electronic degrees of freedom determined by the number of basis functions, $N_\mathrm{B}$, in which the Kohn-Sham orbitals are expanded. Some quantities are vectors in one space and matrices in the other. In general, we do not distinguish both by different types of vector or matrix symbols for different types of spaces, but depending on the context one space is emphasized over the other by the relevant vector or matrix symbol.
    \subsection{General Concept} \label{sec:concept}
        \begin{figure}
            \centering
            \resizebox{\textwidth}{!}{
                \begin{tikzpicture}[node distance=1.5cm]
                    \node (start) [startstop] {Ground-state DFT calculation};
                    \node (in1) [io, below of=start] {Additional ground-state properties};
                    \node (pro1) [process, below of=in1] {Generate $V_{\mathrm{ext}}^{(1)\mathbf{q}}(\mathbf{r})$};
                    \node (pro2) [process, below of=pro1, xshift=5cm] {Set up $\matr{H}^{(1)\mathbf{q}}$ and $\matr{S}^{(1)\mathbf{q}}$};
                    \node (pro3) [process, below of=pro2] {Solve Sternheimer equation};
                    \node (pro4) [process, below of=pro3, yshift=-1cm] {Synthesize $n^{(1)\mathbf{q}}(\mathbf{r})$};
                    \node (dec1) [decision, below of=pro1, yshift=-4cm] {$n^{(1)\mathbf{q}}(\mathbf{r})$ converged?};
                    \node (pro5) [process, left of=pro2, xshift=-8cm] {Generate $V_{\mathrm{eff}}^{(1)\mathbf{q}}(\mathbf{r})$};
                    \node (pro6) [process, below of=dec1, yshift=-2.0cm] {Final iteration for converged quantities};
                    \node (pro7) [process, below of=pro6] {Calculate row of dynamical matrix $\mathbf{D}_{j}^{\beta \mathbf{q}\top}$};
                    \node (pro8) [process, below of=pro7, yshift=-1.5cm] {Evaluate phonon properties};
                    \node(pro9) [process, left of=pro8,xshift=-2.25cm,yshift=1.5cm] {Generate $\matr{V}_{\mathrm{ext}}^{~(2)\mathbf{q}}(\mathbf{r}),\matr{E}_{\mathrm{ion-ion}}^{~(2)\mathbf{q}}$};
                    \draw[draw=red,ultra thick] (-6.5,-3.75) rectangle ++(14.3,-7.15);
                    \draw[draw=violet,ultra thick] (2.2,-3.875) rectangle ++(5.5,-3.4);
                    \node[draw=violet] at (7.2,-6.9) {$\forall\mathbf{k}$};
                    \draw[draw=cyan,ultra thick] (-6.75,-2.3) rectangle ++(14.8,-12.0);
                    \node[draw=cyan] at (7.4,-13.8) {$\forall\beta j$};
                    \draw[draw=blue,ultra thick] (-7.0,-2.15) rectangle ++(15.3,-15.1);
                    \node[draw=blue] at (7.75,-16.75) {$\forall\mathbf{q}$};
                    
                    \draw [arrow] (start) -- node[anchor=west] {Build gradients etc.} (in1);
                    \draw [arrow] (in1) -- (pro1);
                    \draw [arrow] (pro1) -| (pro2);
                    \draw [arrow] (pro2) -- (pro3);
                    \draw [arrow] (pro3) -- (pro4);
                    \draw [arrow] (pro4) -- (dec1);
                    \draw [arrow] (dec1) -| node[anchor=east] {No} (pro5);
                    \draw [arrow] (pro5) -- (pro2);
                    \draw [arrow] (in1) -| (pro5);
                    \draw [arrow] (dec1) -- node[anchor=west] {Yes} (pro6);
                    \draw [arrow] (pro6) -- (pro7);
                    \draw [arrow] (pro7) -- (pro8);
                    \draw [arrow] (pro9.west) -- +(-0.31,0)|- (pro7) node[currarrow,pos=0.4,sloped] {};
                    \draw [arrow] (pro1.west) -- +(-4.85,0) |- (pro7) node[currarrow,pos=0.485,sloped] {};
                \end{tikzpicture}
            }
        \caption{Sketch of workflow for the DFPT calculation leading to phonon properties. The colored frames highlight the loop structure of the calculation. $V_{\mathrm{ext}}^{(1)\mathbf{q}}(\mathbf{r})$, $V_{\mathrm{eff}}^{(1)\mathbf{q}}(\mathbf{r})$, $n^{(1)\mathbf{q}}(\mathbf{r})$, $\matr{H}^{(1)\mathbf{q}}$ and $\matr{S}^{(1)\mathbf{q}}$ are the $(\beta j)$-atom-displacement coordinate  components of the respective vector quantities in the space of the dynamical degrees of freedom.}
            \label{fig:scf1}
    \end{figure}
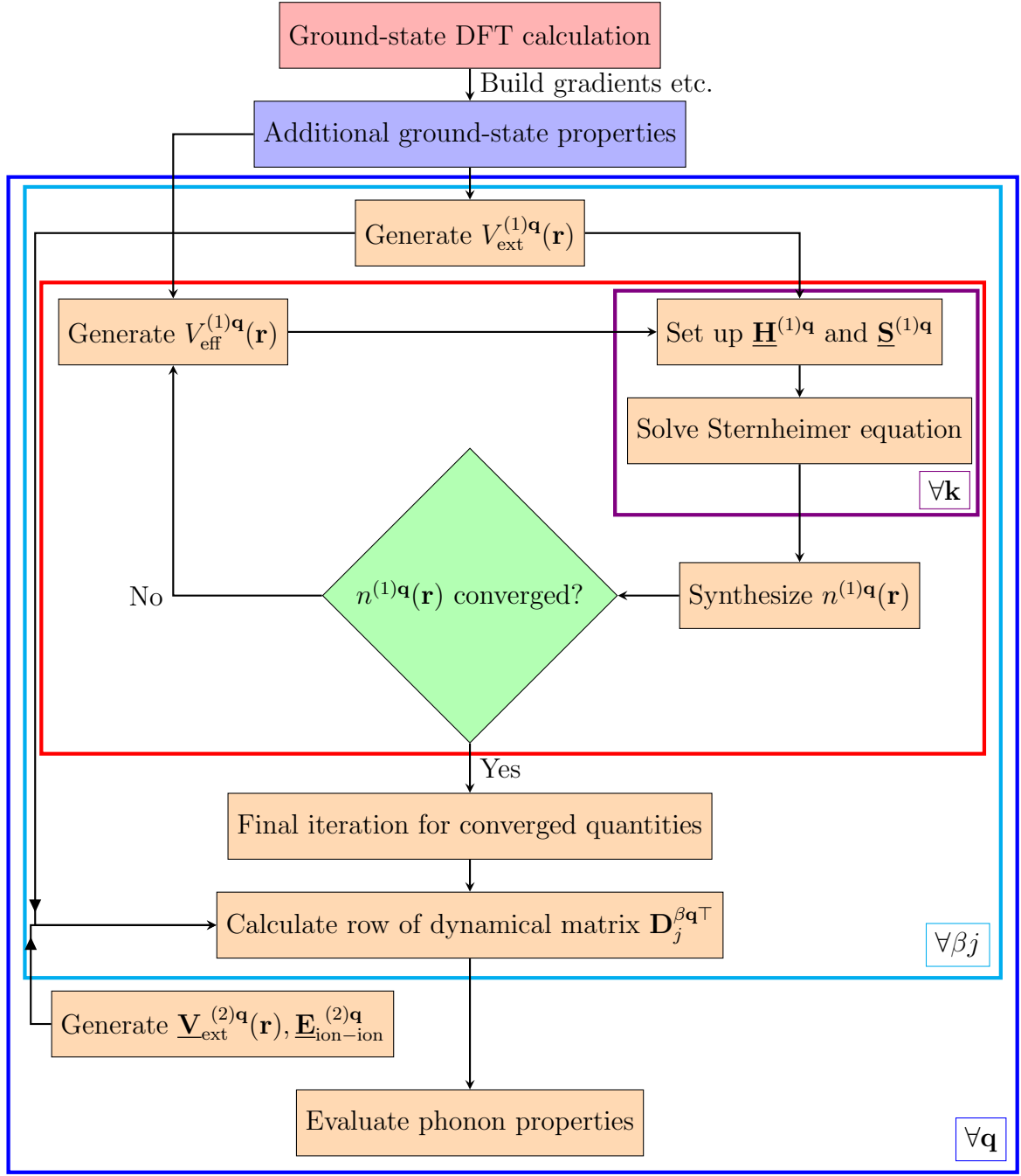
    \noindent  The DFPT formalism to phonon properties consists of three parts: (i) a single ground-state DFT calculation, (ii) the setup and convergence of the Sternheimer equation to obtain the first-order density response $\mathbf{n}^{(1)}$ upon the displacement of atoms, as well as (iii) the setup and diagonalization of the dynamical matrix. Since the phonon is a wavevector $\mathbf{q}$ dependent displacive perturbation, many response or perturbed quantities carry naturally the superscript $\mathbf{q}$ (for more details see section~\ref{ssec:Definitions}). The dynamical matrix is the central input to any phonon property calculator. It is set-up and diagonalized q-vector by q-vector in a totally sequential but parallelizable fashion. The basic  algorithm is sketched in Figure~\ref{fig:scf1}. After the ground-state DFT calculation (red box) was carried out,  we start the DFPT part. For this we need to set-up some additional quantities (blue box) beyond those we determined during the ground state run. These are calculated once before the start of the outer q-loop (blue frame), as they are not q-dependent. They consist of (i) the gradients of the ground-state density and all constituents of the potential (external, Coulomb, xc), as they are needed to determine the basis set corrections and the corrections of the discontinuities at the surface of the MT sphere, and (ii) of the complete set of eigenvalues and eigenvectors of the unperturbed Hamiltonian, which are needed for the quasi-analytical inversion of the Hamiltonian and overlap matrices in the Sternheimer equation. This yields a set of self-consistency equations (red frame) for each q-point, as well as atom in the unit cell and displacement direction (cyan frame). By treating each of the $3N_\mathrm{A}$ perturbations sequentially for each atom $\beta$ and cartesian displacement coordinate $j$, the response quantities are just scalar components of vectorial quantities of dimension $3N_\mathrm{A}$. To keep the overview, we have omitted the indices $\beta, j$ in workflow Figure~\ref{fig:scf1}. We construct the $(\beta j)$-components of the first-order external potential ($\mathbf{V}_{\mathrm{ext}}^{(1)\mathbf{q}}$) as our initial perturbation and establish then the Hamiltonian response ($\matr{H}^{(1)\mathbf{q}}$) together with the overlap matrix ($\matr{S}^{(1)\mathbf{q}}$). From the Sternheimer equation we thus determine for a given k-vector the response of the wave function expansion coefficients for all electronic eigenstates $\nu$. After the k-point loop is completed,  these enter the charge density generator to construct the $(\beta j)$-component of the density response ($\mathbf{n}^{(1)\mathbf{q}}$), which is then used to construct the $(\beta j)$-component of the effective potential response ($\mathbf{V}_{\mathrm{eff}}^{(1)\mathbf{q}}$). This accounts for the self-consistent nature of the problem, since the effective potential response in turn requires a new Hamiltonian and overlap matrix response. We repeat the calculation until the response-density changes less than a given threshold between iterations (red frame). After convergence is reached, a final iteration of the loop is started to construct additional quantities needed to compute the ($\beta j$)-row of the dynamic matrix ($\mathbf{D}_{j}^{\beta \mathbf{q}\top}$). Finally, for this we also need the second-order variations of the external potential $(\matr{V}_{\mathrm{ext}}^{~(2)\mathbf{q}})$ and the ion-ion interaction $(\matr{E}_{\mathrm{ion-ion}}^{~(2)\mathbf{q}})$.
    
    The individual parts of the workflow are laid out in section~\ref{sec:sternheimer}. We deal first with the density response (section~\ref{sec:rho1}), then with the exact form of the Sternheimer equation and the related matrices (section~\ref{sec:sternheimersetup}), the generation of the potential responses (section~\ref{sec:v1}), and finally with the mixing procedure (section~\ref{sec:mixing}). The dynamical matrix setup is found in the succeeding section~\ref{sec:dynmat}. In all calculation steps apart from the mixing, the first and second order quantities are determined in part by quantities obtained from the ground-state calculation, which will be highlighted accordingly.
    \subsection{Definitions}\label{ssec:Definitions}
    First we define the perturbed quantities: provided the periodic displacement $\mathrm{\Delta}\boldsymbol{\tau}$ of atom $\alpha$ in unit cell $\mathbf{R}$ with amplitude $\mathbf{Q}_{\alpha}$ and phonon wave vector $\mathbf{q}$ offsetting the atom from the equilibrium position $\mathbf{R}+\boldsymbol{\tau}_\alpha $ is expressed as
    \begin{IEEEeqnarray}{rCl}\label{eq:DTau}
        \mathrm{\Delta}\boldsymbol{\tau}_{\alpha\mathbf{R}}(\mathbf{q}) = \mathbf{Q}_{\alpha}(\mathbf{q})\mathrm{e}^{\mathrm{i}\mathbf{q}\cdot \mathbf{R}} + \mathbf{Q}_{\alpha}^{*}(\mathbf{q})\mathrm{e}^{-\mathrm{i}\mathbf{q}\cdot \mathbf{R}} \,,
    \end{IEEEeqnarray}
    any perturbed or response quantity $X^{(n)}$ of order $n$, \textit{e.g.}\ think of the charge density response $n^{(1)}$, 
    will be a sum of terms with $n$ different $\pm\mathbf{q}$-dependent phases.
    So, properties originally periodic according to the translation symmetry of the crystalline lattice now carry an additional plane-wave factor, potentially altering the Bloch character.
    For first and second order quantities, $ X^{(1)}$ and  $X^{(2)}$, respectively,  of the general quantity $X$, we subsume  the $N_\mathrm{A}$ atoms ($\alpha$) and the three Cartesian coordinates ($i$) into $3N_\mathrm{A}$ dimensional vectors, $\mathbf{X}^{(1)}$, and $3N_\mathrm{A}\times 3N_\mathrm{A}$ dimensional matrices (underlined quantities), $\matr{X}^{(2)}$, respectively:
    \begin{IEEEeqnarray}{rCl}
        X^{(1)}(\mathbf{q}) &=& \sum_{\alpha,i}Q_{\alpha,i}X^{(1)\alpha i\mathbf{q}}+ Q_{\alpha,i}^{*}X^{(1)\alpha i\mathbf{-q}}\equiv \sum_{\pm}\mathbf{Q}_{\pm}^{\top}\mathbf{X}^{(1)\pm\mathbf{q}}, \\
        X^{(2)}(\mathbf{q}) &=& \frac{1}{2}\sum_{\pm'\pm}\mathbf{Q}_{\pm'}^{\top}\matr{X}^{(2)\pm'\mathbf{q}\pm\mathbf{q}}\mathbf{Q}_{\pm} \, ,
    \end{IEEEeqnarray}
    where superscript $\top$ stands for the transpose operation, and use of the relation $\mathbf{Q}(-\mathbf{q})=\mathbf{Q}^\ast(\mathbf{q})$ was made. The following notation was introduced: Response quantities, either represented as vector or matrices or with indices, represent the direct derivatives with respect to the phononic perturbation, as \textit{e.g.}\ the Hesse matrix to second order, while a absence of indices corresponds to the full scalar perturbation of that order, \textit{i.e.}\ for the same example the contraction of the Hesse matrix with the displacement vectors. $X^{(\cdot)}(\mathbf{q})$ refers the Fourier transform of $X^{(\cdot)}$ over the unit cells.
    Working with these projections enables us to suppress the dimensional character of the involved quantities.  
    It can easily be shown, by requiring integrals over the unit cell to be non-vanishing, that (i) to second order only combinations of $+-$ or $-+$ contribute~\cite{Savrasov1996_PhysRevB_54_16470} and (ii) those terms are the Hermitian conjugates of each other.
    To that end, it is sufficient to solely calculate the $+$ part to first order in the eigenstates, charge density, and potential perturbations as well as to only calculate the DM for the combination $+-$. 
    By denoting the atomic and directional indices for the columns by $\alpha i$ and the rows by $\beta j$, we need to calculate quantities such as:
    \begin{IEEEeqnarray}{rCl}
        n^{(1)\beta j+}(\mathbf{q}) &\coloneq& \sum_{\mathbf{R}}\mathrm{e}^{\mathrm{i}\mathbf{q}\cdot \mathbf{R}}\frac{\partial n}{\partial \tau_{\beta\mathbf{R}j}}|_{\boldsymbol{\tau}=0}\,,\label{eq:comp_not}\\
        V_{\mathrm{eff}}^{(1)\beta j+}(\mathbf{q}) &\coloneq& \sum_{\mathbf{R}}\mathrm{e}^{\mathrm{i}\mathbf{q}\cdot \mathbf{R}}\frac{\partial V_{\mathrm{eff}}}{\partial \tau_{\beta\mathbf{R}j}}|_{\boldsymbol{\tau}=0}\,,\\
        E_{\mathrm{tot}}^{(2)\beta j+\alpha i-}(\mathbf{q}) &\coloneq& \sum_{\mathbf{R'},\mathbf{R}}\mathrm{e}^{\mathrm{i}\mathbf{q}\cdot( \mathbf{R'}-\mathbf{R})}\frac{\partial^{2} E_{\mathrm{tot}}}{\partial \tau_{\beta\mathbf{R'}j}\partial \tau_{\alpha\mathbf{R}i}}|_{\boldsymbol{\tau}=0} \,. \IEEEeqnarraynumspace
    \end{IEEEeqnarray}
    
    \subsection{Challenges} \label{sec:Challenges}  
       The LAPW basis function adds a position-dependent basis set in the MT region to the basis consisting of plane waves in the IR, which leads to non-trivial additional complexities. 
       When the wave function is varied due to a phonon perturbation, owing to the displacement sensitivity of the basis, the first-order change of the wave function is not expressed only by the first-order change of the expansion coefficients to a momentum increased by the phonon wave vector $\mathbf{q}$, $z_{\mathbf{k}+\mathbf{G}+\mathbf{q},\nu}^{(1)}$, 
        \begin{IEEEeqnarray}{rCl}
            \Psi^{(1)}_{\mathbf{k}+\mathbf{q},\nu}(\mathbf{r})  &=& \sum_{\mathbf{G}} z_{\mathbf{k}+\mathbf{G}+\mathbf{q},\nu}^{(1)} \phi_{\mathbf{k}+\mathbf{G}+\mathbf{q}}(\mathbf{r}) + \sum_{\mathbf{R}}\mathrm{e}^{\mathrm{i}\mathbf{q}\cdot \mathbf{R}}\sum_{\mathbf{G}} z_{\mathbf{k}+\mathbf{G},\nu} \phi^{(1)\mathbf{R}}_{\mathbf{k}+\mathbf{G}}(\mathbf{r}) \IEEEnonumber \\
            &=& \Psi^{(1),\in \mathrm{HS}}_{\mathbf{k}+\mathbf{q},\nu} + \Psi^{(1),\notin \mathrm{HS}}_{\mathbf{k}+\mathbf{q},\nu} ,\label{eq:psi_one_HS}
        \end{IEEEeqnarray}
        but also by a term that is the derivative $\phi^{(1)\mathbf{R}}_{\mathbf{k}+\mathbf{G}}(\mathbf{r})$ of the basis function with respect to the atomic displacement in a unit cell $\mathbf{R}$. This latter term (and all related quantities) lies typically outside the Hilbert space (HS) spanned by the original LAPW basis functions and cannot be included efficiently by increasing the number of functions.
        We distinguish between two out-of-HS contributions: (i) the Pulay terms~\cite{Pulay1969}, akin to a set of corrections in atomic-force calculations~\cite{Yu1991_PhysRevB_43_6411}, 
        and (ii) the variation of the effective potential, affecting the radial solutions $u_\ell$ in the MT part of the LAPW basis (basis response)~\cite{PhysRevB.85.245124}.
        Both terms together are known as the incomplete basis set correction (IBC).
        While the Pulay terms are indispensable for a successful DFPT calculation in the FLAPW method, the basis response is (according to literature for force calculations) assumed to be small~\cite{Yu1991_PhysRevB_43_6411}. This results in the so-called frozen-augmentation approximation,
        where this contribution is neglected and that we likewise adopt. The remaining terms stem from the differentiation of the matching coefficients, yielding an imaginary prefactor, and the direct differentiation of the position dependence, which is expressed as a gradient, $\boldsymbol{\nabla}$, with respect to the space coordinate  $\mathbf{r}$. Written with the full set of response indices, we find 
        \begin{IEEEeqnarray}{rCl}
        \phi^{(1)\beta\mathbf{R}j}_{\mathbf{k}+\mathbf{G}}(\mathbf{r})=\mathrm{\Theta}_{\beta\mathbf{R}} (\mathbf{r})\left(\mathrm{i} \left(\mathbf{k} + \mathbf{G}\right)_{\!j}-\nabla_{\!\!j}\right)\phi_{\mathbf{k}+\mathbf{G}}(\mathbf{r})\, . 
        \end{IEEEeqnarray}
        The latter gradient operator poses a numerical challenge, as we have to calculate the corresponding radial derivative of $u_\ell$ on a finite radial grid that extends to the atomic nucleus, where this can become inaccurate. 
        Another challenge lies in the numerical discontinuity of the LAPW basis functions at the MT boundary (see also discussion in \ref{sec:app_ekin}).
        Although the LAPW basis is in principle differentiable throughout the unit cell when the real-space basis function in the MT region is expanded into an unlimited number of angular momentum coefficients $\ell$, reasonable cutoffs lead to slight discontinuities in zeroth and first order or even severe discontinuities in higher orders.
        Especially in the case of phonons, these higher orders become relevant (\textit{e.g.}\ when applying a dyadic product
        of gradients to the basis function).
        To mitigate this and the dependence of integrals over quantities $X$ involving wave functions on the positions of the moving nuclei,
        additional surface integrals need to be considered~\cite{KlueppelbergDr}:
        \begin{IEEEeqnarray}{rCl}
        \frac{\partial}{\partial\tau_{\beta\mathbf{R}j}}\int_{\mathrm{\Omega}}X(\mathbf{r})\mathrm{d}\mathbf{r}=\int_{\mathrm{\Omega}}\frac{\partial X}{\partial\tau_{\beta\mathbf{R}j}}(\mathbf{r})\mathrm{d}\mathbf{r}+\oint_{\partial \beta\mathbf{R}} [X]_{\mathrm{SF}}\uv{e}_{j}\mathrm{d}S.\label{eqn:basicsurface}
        \end{IEEEeqnarray}
        Terms of the form $[X]_{\mathrm{SF}}$ are to be understood as the difference of the function taken in the MT domain ($\partial \beta^+$) and in the interstitial domain ($\partial \beta^-$). 
        The former can be used to cancel one problem against another. In practice, each response of a $\mathbf{r}$-dependent quantity has a certain resemblance to the gradient of the same quantity in the displaced MT-sphere. For the response of the basis function, this similarity is analytically explicit, while for quantities like the potential responses, it is implicit. Overall, it is beneficial to sum up response quantities with the corresponding gradient. The gradients can be readily obtained at various points in the calculation, due to the fact that any surface integral of a function over a closed surface can be rewritten as an integral over the enclosed volume of the function's gradient. We make heavy use of this and hence eliminate the necessity to deal with gradient terms as often as possible, which leads to the regrouping of terms and the existence of surface integrals of IR quantities, that are not paired with their corresponding MT representation anymore.

        As a last aside, we need a numerically stable representation of the various Coulomb potentials and their derivatives, as well as the Coulomb energy between the nuclei, despite the $1/r~$-singularities at the nuclei, in particular when a gradient or a dyadic product of gradients is involved. Here the method of Weinert~\cite{Weinert1981_JMathPhys_22.2433} for solving Poisson's equation without shape approximation for an arbitrary periodic charge distribution can be used and generalized. The Coulomb terms are then perfectly continuous by construction (\textit{i.e.}\  chapter~\ref{sec:v1}). 
        It turns out that the standard \texttt{FLEUR} integration scheme (6-point Simpson integrator) for radial integrals in the MT sphere has trouble handling the density gradient and related quantities, due to wild oscillations at the core that propagate outwards, and it was thus replaced by a 4-point spline integrator in the DFPT part of the code.

    \subsection{Sternheimer Equation} \label{sec:sternheimer}
        We start with elaborating on the first-order density variation, then discuss the setup of the Sternheimer equation, shortly introduce how we calculate the linear potential variations and close with a brief overview of our strategy to achieve self-consistency in the Sternheimer equation.
        \subsubsection{Linear Density Response} \label{sec:rho1}
            The representation of the first-order density response $n^{(1)}$ around the unperturbed density $n$ in the FLAPW method comprises various terms. Ultimately, it comes back to the wave function response.
            The wave function part related to the temperature dependent occupation function ($\tilde{f}_{\mathbf{k}\nu}$) and expansion coefficient ($z_{\mathbf{k}+\mathbf{G},\nu}$) responses are very similar to the plane-wave part in the pseudopotential method~\cite{Baroni2001_RevModPhys_73_515}, 
            but with the basis-function response we instead find (using the simplification of~\ref{sec:app_psi1})
            \begin{IEEEeqnarray}{rCl}
             n^{(1)\beta j+}(\mathbf{r})
                =&\hphantom{+}\sum_{\mathbf{k}\nu} \tilde{f}_{\mathbf{k}\nu} \Psi_{\mathbf{k}\nu}^{*}(\mathbf{r})\times 2\sum_{\mathbf{G}}&{} \left(z_{\mathbf{k}+\mathbf{G}+\mathbf{q},\nu}^{(1)\beta j}\phi_{\mathbf{k}+\mathbf{G}+\mathbf{q}}(\mathbf{r})\right.\IEEEnonumber\\
                &{}&\left.+ \mathrm{\Theta}_{\beta} (\mathbf{r})\left(\mathrm{i} \left(\mathbf{k} + \mathbf{G}\right)_{\!j}-\nabla_{\!\!j}\right)z_{\mathbf{k}+\mathbf{G},\nu} \phi_{\mathbf{k}+\mathbf{G}}(\mathbf{r})\right)\IEEEnonumber\\
                &+\sum_{\mathbf{k}\nu} \tilde{f}_{\mathbf{k}\nu}^{(1)\beta j}|\Psi_{\mathbf{k}\nu}(\mathbf{r})|^{2},~~~~&\label{eqn:rho1GeneralFLAPW2}
            \end{IEEEeqnarray}
            where a possible spin index is omitted again.
            Explicit terms are induced as (i) the "direct" response of the expansion coefficients, (ii) the consequence of the basis set variation that creates a term with an imaginary prefactor in the displaced MT sphere and a gradient term (that is in practice grouped with its complex conjugate into a $\nabla_{\!\!j}n$-term), as well as (iii) a term dependent on the perturbed occupation function of the electronic state (for materials like metals where the occupation is fractional).
            Thus, we need to precalculate the gradient of the unperturbed density (as well as gradients of the potential for steps later in the calculation). The corresponding formulae and details on the numerical accuracy can be found in reference~\cite{GerhorstDr:}. It is noteworthy to realize that the density response field is a functional of the unperturbed density, its gradient, the first order change of the expansion coefficient, which is a functional of the first order change of the potential, which depends on the first order change of density, $n^{(1)}[n, \nabla n, z^{(1)}[V^{(1)}[n, n^{(1)}]]]$.
            
            Evaluating~\eqref{eqn:rho1GeneralFLAPW2} for the part independent of $\tilde{f}_{\mathbf{k}\nu}^{(1)\beta j}$ in the IR results in
            \begin{IEEEeqnarray}{rCl}
                n_{\mathrm{IR},\tilde{f}}^{(1)\beta j+}(\mathbf{r})
                \hspace{-0cm}&=& \frac{2}{\mathrm{\Omega}} \sum_{\mathbf{k}\nu} \tilde{f}_{\mathbf{k}\nu} \sum_{\mathbf{G}''} \sum_{\mathbf{G}'} z_{\mathbf{k}+\mathbf{G}',\nu}^{*} z_{\mathbf{k}+\mathbf{G}''+ \mathbf{q},\nu}^{(1)\beta j} \mathrm{e}^{\mathrm{i} \left(\mathbf{G}'' - \mathbf{G}' + \mathbf{q} \right) \cdot \mathbf{r}} \IEEEyesnumber\IEEEyessubnumber \\
                &=& \sum_{\mathbf{G},\mathbf{G}+\mathbf{q}\neq0} n_{\mathrm{IR}}^{(1)\beta j}(\mathbf{G}+\mathbf{q}) \mathrm{e}^{\mathrm{i} \left(\mathbf{G} + \mathbf{q}\right) \cdot \mathbf{r}} \IEEEyessubnumber \label{eqn:implStHmDens1AlgIRFinLinRunOneSum} \,,
            \end{IEEEeqnarray}
            whereas in the MT sphere of atom $\gamma$ we implement
            \begin{IEEEeqnarray}{rCl}
                n_{\mathrm{MT}^{\gamma}}^{(1)\beta j+}(\mathbf{r}) &=& \sum_{L} n_{\mathrm{MT}^{\gamma},L}^{(1)\beta j}(r_{\gamma})Y_{L}(\mathrm{\hat{r}}_\gamma) \IEEEyesnumber\IEEEyessubnumber \\
                n_{\mathrm{MT}^{\gamma},L}^{(1)\beta j}(r_{\gamma})
                &=&  \sum_{\ell' p' \ell'' p''}
                u^{\gamma}_{\ell' p'}(r_{\gamma})
                u^{\gamma}_{\ell'' p''}(r_{\gamma})d_{\ell' p' \ell'' p''}^{\gamma,\beta j}(L)-\mathrm{\delta}_{\gamma\beta}\left[\nabla_{j} n\right]_{\!\mathrm{MT}^{\gamma},L}(r_{\gamma})\IEEEyessubnumber \label{eqn:rho1MTcoefficients} \, \IEEEeqnarraynumspace,
            \end{IEEEeqnarray}
            where the brackets $[\,\cdot\,]$ in $\left[\boldsymbol{\nabla} n\right]_{\mathrm{MT}^{\gamma},L}(r_{\gamma})$ denote the expansion of the density gradient into  lattice harmonics~\cite{Altmann:65}. The coefficients $d$, containing the linear response of the matching coefficients, expansion coefficients, and occupation numbers, are defined in~\ref{sec:app_rho1}.
            Within these interstitial and muffin-tin representations, the central quantity not given by a preceding DFT calculation is the first-order variation of the wave function coefficients $z_{\mathbf{k}+\mathbf{G}+\mathbf{q},\nu}^{(1)\beta j}$. 
            They are determined by a self-consistent solution of the Sternheimer equation.
            
            \textsl{Core electrons} In the charge density response quantity, $n^{(1)}$, we explicitly consider the valence states only, \textit{i.e.}\ we apply the frozen-core approximation, stating the core electrons not to be perturbed by a shift of the atomic positions, while in the ground state density and the gradient of the ground state density, $\boldsymbol{\nabla}n$, the full density enters, \textit{i.e.}\ including the core electrons.
            In \texttt{FLEUR}, there is furthermore an option to explicitly consider electrons leaking out of the MT spheres --- they originate from and permeate other muffin-tins --- by applying so-called core-tail corrections~\cite{singhnordstrom,PhysRevB.32.7792}, which would warrant additional terms to first order. 
            We postpone the implementation of their perturbation~\cite{PhysRevB.91.035105}, because core-tails can be suppressed by using local orbitals~\cite{singh}.
            The core electrons do, however, contribute to the gradient of the all-electron charge density in the same equation.
        \subsubsection{Setup of the Sternheimer Equation} \label{sec:sternheimersetup}
            As mentioned before, the Sternheimer equation takes a more lengthy form in LAPW methods as compared to pure plane-wave formulations.
            Inserting~\eqref{eq:psi_one_HS} into~\eqref{eqn:verysimpleSH} and explicitly accounting for the basis variation and the surface terms yields in the space of LAPW basis functions
            \begin{IEEEeqnarray}{rCl}
                &&\sum_{\mathbf{G}}  \Braket{\phi_{\mathbf{k} + \mathbf{G}' + \mathbf{q}} | \mathscr{H}_{\vphantom{\mathbf{k}\nu}} - \epsilon_{\mathbf{k}\nu} | \phi_{\mathbf{k}+\mathbf{G}+\mathbf{q}}}_{\!\!\!~\mathrm{\Omega}} z_{\mathbf{k}+\mathbf{G}+\mathbf{q},\nu}^{(1)\beta j} \IEEEnonumber
                \\
                = - &&\sum_{\mathbf{G}}  \left\{ \hspace{-.2cm}\vphantom{\oint_{\partial \alpha}} \right. \Braket{\phi_{\mathbf{k} + \mathbf{G}' + \mathbf{q}} | {V_{\mathrm{eff}}^{(1)\beta j+}} - \mathrm{\delta}_{\mathbf{q}, \mathbf{0}} \epsilon_{\mathbf{k}\nu}^{(1)\beta j}| \phi_{\mathbf{k}+\mathbf{G}}}_{\!\!\!~\mathrm{\Omega}}  \IEEEnonumber  \\
                &&{}+ \Braket{\phi^{(1) \beta  j-}_{\mathbf{k} + \mathbf{G}' + \mathbf{q}} | \mathscr{H}_{\vphantom{\mathbf{k}\nu}} - \epsilon_{\mathbf{k}\nu} | \phi_{\mathbf{k}+\mathbf{G}}}_{\!\!\!~\beta} + \Braket{\phi_{\mathbf{k} + \mathbf{G}' + \mathbf{q}} | \mathscr{H}_{\vphantom{\mathbf{k}\nu}} - \epsilon_{\mathbf{k}\nu} | \boldsymbol{\phi}^{(1) \beta j +}_{\mathbf{k}+\mathbf{G}}}_{\!\!\!~\beta}  \label{eqn:SternheimerBasicFLAPW}\\
                &&{}+ \left. \oint_{\partial \beta} \left[\phi^{*}_{\mathbf{k} + \mathbf{G}' + \mathbf{q}}(\mathbf{r}) \left(\mathscr{H}_{\vphantom{\mathbf{k}\nu}} - \epsilon_{\mathbf{k}\nu}\right) \phi_{\mathbf{k}+\mathbf{G}}(\mathbf{r})\right]_{\!\mathrm{SF}} \uv{e}_{j}\mathrm{d}S \right\}z_{\mathbf{k}+\mathbf{G},\nu} \IEEEnonumber \, . 
            \end{IEEEeqnarray}
           The first line constitutes the Hellmann--Feynman contribution, the second line contains Pulay terms, and the third line contains surface terms. In this representation, the Pulay terms are of significant value. They consist of a prefactor part and a part containing the gradient of the basis function. The latter are not so numerically well-behaved, especially in the core region, that their numerical integration over the muffin-tin sphere guarantees sufficient accuracy for reliable phonon properties.
           
           To solve this equation, we exploit that the left side is a matrix-vector product, and we need to invert the matrix. Instead of working in the space spanned by the LAPW basis functions, we switch to a representation of the Sternheimer equation where the space is spanned by the Kohn--Sham wave functions by multiplying~\eqref{eqn:SternheimerBasicFLAPW} from the left with $z_{\mathbf{k}+\mathbf{G}'+\mathbf{q},\nu}^{*}$ and contracting over $\mathbf{G}'$.
            This procedure avoids a costly inversion of a matrix that is nearly singular at certain eigenvalues.
            Given the definitions of both the band representation of the perturbed expansion coefficients
            \begin{IEEEeqnarray}{rCl}  z_{\mathbf{k}+\mathbf{G}+\mathbf{q},\nu}^{(1)\beta j} &\equiv& \sum_{\nu'} z_{\mathbf{k}+\mathbf{G}+\mathbf{q},\nu'} z_{\mathbf{q}\nu',\mathbf{k}\nu}^{(1)\beta j}\label{eq:z1fromband}
            \end{IEEEeqnarray} 
            and the prefactor part of the perturbed wave function
            \begin{IEEEeqnarray}{rCl}
                \widetilde{\Psi}_{\mathbf{k} + \mathbf{q},\nu}^{j} \coloneq  \mathrm{i} \sum_{\mathbf{G'}} \left(\mathbf{k} + \mathbf{G}' + \mathbf{q}\right)_{j} z_{\mathbf{k}+\mathbf{G}'+\mathbf{q},\nu} \phi_{\mathbf{k} + \mathbf{G}' + \mathbf{q}} \, , \IEEEyesnumber
            \end{IEEEeqnarray}
     we rewrite~\eqref{eqn:SternheimerBasicFLAPW} into
            \begin{IEEEeqnarray}{rCl}
                z_{\mathbf{q}\nu',\mathbf{k}\nu}^{(1)\beta j}
                &=&-\frac{1}{\epsilon_{\mathbf{k} + \mathbf{q},\nu'} - \epsilon_{\mathbf{k}\nu}}\left\{ \vphantom{\oint_{{\partial \mathrm{MT}}(\beta, \mathbf{0})}} \right.
                \Braket{\Psi_{\mathbf{k} + \mathbf{q},\nu'} | 
                (V_{\mathrm{eff}}\mathrm{\Theta}_{\mathrm{IR}})^{(1)\beta j+} |
                \Psi_{\mathbf{k}\nu}}_{\!\!\!~\mathrm{IR}} \IEEEnonumber \\
                &&{}+ \sum_{\gamma} \Braket{\Psi_{\mathbf{k} + \mathbf{q},\nu'} |
                V_{\mathrm{eff}}^{(1)\beta j+} + \mathrm{\delta}_{\gamma \beta}\nabla_{\!\!j} V_{\mathrm{eff}} |
                \Psi_{\mathbf{k}\nu}}_{\!\!\!~\gamma}
                \IEEEnonumber \\
                &&{}+ \Braket{\widetilde{\Psi}_{\mathbf{k} + \mathbf{q},\nu'}^{j} |
                \mathscr{H_{\vphantom{\mathbf{k}\nu}}} - \epsilon_{\mathbf{k}\nu} |
                \Psi_{\mathbf{k}\nu}}_{\!\!\!\!~\beta} + \Braket{\Psi_{\mathbf{k} + \mathbf{q},\nu'} |
                \mathscr{H_{\vphantom{\mathbf{k}\nu}}} - \epsilon_{\mathbf{k}\nu} |
                \widetilde{\Psi}_{\mathbf{k}\nu}^{j}}_{\!\!\!\!~\beta}
                \label{eqn:SternheimerEquationImplemented} \\
                &&{}+ \left.\Braket{\Psi_{\mathbf{k} + \mathbf{q},\nu'} | 
                \mathrm{\Theta}_{\mathrm{IR}}^{(1)\beta j+}(\mathscr{T}_{\vphantom{\mathbf{k}\nu}} - \epsilon_{\mathbf{k}\nu}) |
                \Psi_{\mathbf{k}\nu}}_{\!\!\!~\mathrm{IR}}  \right\},
                \IEEEnonumber 
            \end{IEEEeqnarray}
            with the kinetic energy operator $\mathscr{T}$ (on which some notes are found in \ref{sec:app_ekin}). Since $V_{\mathrm{eff}}^{(1)}$ depends on $n^{(1)}$ and, therefore, on $z^{(1)}$, this Sternheimer equation must be solved self-consistently according to the scheme in Figure~\ref{fig:scf1}.
            Comparing this form with the initial Sternheimer equation~\eqref{eqn:SternheimerBasicFLAPW}, which subdivides into the Hellmann--Feynman (first line), the Pulay~(2nd line) and the surface terms (last line), we now group the contributions differently (as discussed in section~\ref{sec:Challenges}). 
            We highlight (i) the complete representation of the Sternheimer equation in the Kohn--Sham wave function spanned Hilbert space (and consequently the contraction of the~$\mathbf{G}$-vectors), (ii) the summation of the first-order effective potential and the gradient of the unperturbed potential in the muffin-tin matrix element of atom $\gamma$, that avoids the integration over large terms around the center of the MT spheres, (iii) the overall avoidance of gradients of wave functions and thus contributions outside the established LAPW Hilbert space by cancelling them with the MT surface terms, and (iv) the grouping of IR terms into a combined perturbation of the interstitial potential and step function 
            \begin{IEEEeqnarray}{rCl}
            \mathrm{\Theta}_{\mathrm{IR}}^{(1)\beta j+} (\mathbf{G}+\mathbf{q})=\mathrm{i}(G+q)_{j}
            \mathrm{\Theta}_{\beta}(\mathbf{G}+\mathbf{q})\, ,
            \end{IEEEeqnarray}
            in Fourier representation.
            Considering (ii), we introduce a shorthand notation for combinations of perturbations and gradients in the displaced MT. We write
            \begin{IEEEeqnarray}{rCl}
                X^{(1)\beta j+\nabla} &=& X^{(1)\beta j+} + \mathrm{\delta}_{\gamma \beta}\nabla_{j} X \,
            \end{IEEEeqnarray}
            to streamline further equations, as such combinations reappear frequently in the dynamical matrix setup.
            Details on the general evaluation of IR or MT matrix elements are pointed out in reference~\cite{GerhorstDr:}.
            It becomes obvious that knowledge about the unperturbed system at shifted Bloch vectors $\mathbf{k}+\mathbf{q}$ is required. 
            We choose to calculate all that information once in the beginning, before the Sternheimer loop of a particular q-point.
            As a test of the implementation, one can show that the analytical solution of the Sternheimer equation for $\mathbf{q}=\mathbf{0}$ and one atom is ~\cite{GerhorstDr:}
            \begin{IEEEeqnarray}{rCl}
                z_{\mathbf{k}+\mathbf{G},\nu}^{(1)j} = - \mathrm{i} \left(\mathbf{\mathbf{k}} + \mathbf{G}\right)_{j} z_{\mathbf{k}+\mathbf{G},\nu}\label{eq:an_sol} \,.
            \end{IEEEeqnarray}
            
            There are two more things to consider. 
            Firstly,~\eqref{eqn:SternheimerEquationImplemented} only holds for non-vanishing energy differences $\delta_{\mathbf{q}\nu',\mathbf{k}\nu}\coloneq|\epsilon_{\mathbf{k} + \mathbf{q},\nu'} - \epsilon_{\mathbf{k}\nu}|$. We group terms with and without a prefactor $\epsilon_{\mathbf{k}\nu}$ together and identify them as a perturbed overlap matrix $\matr{S}^{(1)}$ and Hamiltonian $\matr{H}^{(1)}$, respectively, also referred to as overlap matrix response and Hamiltonian response in the space of the $N_\mathrm{B}\times N_\mathrm{B}$ Kohn-Sham states and at the same time $3N_\mathrm{A}$ dimensional vectors in the space spanning the dynamical matrix,  to rewrite~\eqref{eqn:SternheimerEquationImplemented} as
            \begin{IEEEeqnarray}{rCl}
                z_{\mathbf{q}\nu',\mathbf{k}\nu}^{(1)\beta j+}\equiv-\frac{1}{\epsilon_{\mathbf{k} + \mathbf{q},\nu'} - \epsilon_{\mathbf{k}\nu}}\left(H_{\mathbf{q}\nu',\mathbf{k}\nu}^{(1)\beta j+}-\epsilon_{\mathbf{k}\nu}S_{\mathbf{q}\nu',\mathbf{k}\nu}^{(1)\beta j+}\right)\IEEEyesnumber \label{eqn:Sternheimershort}\, .
            \end{IEEEeqnarray}
            If the energy difference $\delta_{\mathbf{q}\nu',\mathbf{k}\nu}$ is close to zero (a threshold of $10^{-12}$~htr was used for the calculations in this work), a special treatment is required in order to avoid an explicit division by very small numbers, and the following reformulated expression (see derivation in~\ref{sec:app_de})
            \begin{IEEEeqnarray}{rCl}
                z_{\mathbf{q}\nu',\mathbf{k}\nu}^{(1)\beta j+} [\delta_{\mathbf{q}\nu',\mathbf{k}\nu}\approx0]
                = &-&\frac{1}{2}S_{\mathbf{q}\nu',\mathbf{k}\nu}^{(1)\beta j+}= z_{\mathbf{q}\nu',\mathbf{k}\nu}^{(1)\beta j+,\mathrm{mod}} \,, \label{eqn:SternheimerEquationModded}
            \end{IEEEeqnarray}
            is applied.
            Secondly, arithmetic and sum reformulations yield an individual procedure for the case in which the energy difference is finite and both $\nu',\nu$ represent occupied states, meaning  their occupation-number prefactor $\tilde{f}$ is larger than a certain threshold (set by default to $10^{-8}$ in FLEUR).
            Provided this condition, we can derive
            \begin{IEEEeqnarray}{rCl}
                z_{\mathbf{q}\nu',\mathbf{k}\nu}^{(1)\beta j+} [\mathrm{occ}-\mathrm{occ}]
                = (1-F(\epsilon_{\mathbf{k}+\mathbf{q},\nu'}))z_{\mathbf{q}\nu',\mathbf{k}\nu}^{(1)\beta j+}-\frac{1}{2}F(\epsilon_{\mathbf{k}+\mathbf{q},\nu'})S_{\mathbf{q}\nu',\mathbf{k}\nu}^{(1)\beta j+}=z_{\mathbf{q}\nu',\mathbf{k}\nu}^{(1)\beta j+,\mathrm{occ}}
                \label{eqn:SternheimerEquationModded2} \,, \IEEEeqnarraynumspace
            \end{IEEEeqnarray}
where $F(\epsilon_{\mathbf{k}+\mathbf{q},\nu'})$ is the Fermi smearing for the respective eigenenergy (see \ref{sec:app_occ}). Its smoothness can be controlled by the Fermi smearing parameter $k_{\mathrm{B}}T$. Using this modification can improve the stability of the self-consistency calculation. 
       
        \subsubsection{Potential Responses} \label{sec:v1}
            $V^{(1)}_\mathrm{eff}$ and $\boldsymbol{\nabla}V_\mathrm{eff}$, the first order response and the gradient of the effective potential $V_\mathrm{eff}$ both enter the Sternheimer equation~\eqref{eqn:SternheimerEquationImplemented} as well as  the set-up of the dynamical matrix. For the latter, we also need the response and gradient of the external $V_\mathrm{ext}$ 
            and of the Coulomb potential $V_\mathrm{C}$, due to various correction terms that occur as a consequence of the LAPW basis (see section~\ref{sec:dynmat}). Since $V_\mathrm{eff}=V_\mathrm{C}+V_\mathrm{xc}=V_\mathrm{H}+V_\mathrm{ext}+V_\mathrm{xc}$, we need each term up to first order and also the corresponding real-space gradients. 
            We briefly discuss the calculations of these terms in the following.
            
            \textsl{Hartree and external potential} -- The Hartree potential response, $V^{(1)}_\mathrm{H}$,  is basically  the Hartree potential of the response charge density $n^{(1)}$, 
            \begin{IEEEeqnarray}{rCl}
            V_{\mathrm{H}}^{(1)\beta j+}(\mathbf{r})&=&\int\frac{n^{(1)\beta j+}(\mathbf{r}')}{\left|\mathbf{r}-\mathbf{r}'\right|}\mathrm{d}\mathbf{r}'+\oint_{\partial \beta} \frac{[n(\mathbf{r}')]_{\mathrm{SF}}}{\left|\mathbf{r}-\mathbf{r}'\right|}\uv{e}_{j}\mathrm{d}S' \, ,
            \label{eqn:vCoul1}
            \end{IEEEeqnarray}
            and the Hartree potential gradient, $\boldsymbol{\nabla} V_{\mathrm{C}}$, is basically the Hartree potential of the gradient of the charge density $\boldsymbol{\nabla} n$,
            \begin{IEEEeqnarray}{rCl}
            \nabla_{j}V_{\mathrm{H}}(\mathbf{r})&=&\int\frac{\nabla_{j}n(\mathbf{r}')}{\left|\mathbf{r}-\mathbf{r}'\right|}\mathrm{d}\mathbf{r}'-\sum_{\gamma}\oint_{\partial \gamma} \frac{[n(\mathbf{r}')]_{\mathrm{SF}}}{\left|\mathbf{r}-\mathbf{r}'\right|}\uv{e}_{j}\mathrm{d}S'   \label{eqn:vgCoul} \, ,
            \end{IEEEeqnarray} plus additional surface integrals introduced in \eqref{eqn:basicsurface}. These surface corrections  apply for the displaced atom in the potential response calculation and for all atoms in the gradient case and correct possible discontinuities at the muffin-tin boundary.  Therefore, we use the Weinert algorithm~\cite{Weinert1981_JMathPhys_22.2433} for solving the Poisson equation to obtain both $V_{\text C}^{(1)}$ and $\nabla_{j} V_{\mathrm{C}}$ from the charge density response to first order and the gradient of the unperturbed charge density, respectively. 
            The procedure is similar to that of the ground-state calculation, where the radially dependent MT densities are replaced by a smooth Fourier transformable pseudo-density valid in the whole unit cell, so that the interstitial potential can be directly expressed by the pseudo-density components, while  the MT potential is obtained by solving  a boundary value problem inside the MT sphere. However, instead of using the density, we use the first-order response density or the gradient of the density. 
            We employ the Weinert algorithm not only for the response, but also for the gradient of the Coulomb potential, since continuity at the muffin-tin boundary (being important for well-behaving numerics) is then ensured by construction, which is not the case if we straight-forwardly differentiate the ground-state Coulomb potential across the sphere boundary. 
            
           Basically, we  employ (with small modifications) the same Coulomb solver routines for the potential response as for the unperturbed potential, but there are some points to consider. Firstly, additional care has to be taken with respect to the radial integration of the density response or density gradient in the Coulomb solver, as they can be less smooth than the typically used ground-state density. A second  point is the emergence of surface terms in \eqref{eqn:vCoul1} and \eqref{eqn:vgCoul}. 
            We express them here as a correction to the basic multipole moments $q_{lm}$ of degree $(l, m)$ as described in~\cite{Weinert1981_JMathPhys_22.2433}. Keeping to Weinert's original notation (Equation (11) of his paper), the effective multipole moment of the response charge density inside the $i$th MT sphere can be written as
            \begin{IEEEeqnarray}{rCl}
                \tilde{q}_{lm}^{i} = q_{lm}^{i}-q_{lm}^{iI} + q_{lm}^{i,\mathrm{SF}}-q_{lm}^{iI,\mathrm{SF}} .
            \end{IEEEeqnarray}
            The first two terms describe the multipole moments of the true response charge density in the sphere of atom $i$ subtracted by the multipole moments of the plane-wave response charge in the ith atom. 
            The second two terms, denoted by the superscript SF, correspond to corrections from the surface integral.  
            Lastly, the infinitesimal displacement of the Coulomb singularity of the atoms contributes in first order response term by an $(\ell=1)$-character instead of being spherical with $(\ell=0,m=0)$.
            Aside from this, the Weinert procedure is used as in his seminal paper. The specifics of the modified terms can be found in~\ref{sec:app_pot}, while the full derivation of the adapted method was elaborated on in reference~\cite{KlueppelbergDr}.
            Like in the original method, where the combination of the Hartree and external potential mitigates the $\sim 1 / r$ singularity, here the $\sim 1/r^2$ contributions with large absolute values compensate each other and lead to a better controllable numerical behaviour.
            
            \textsl{Exchange correlation potential} -- In order to calculate the first-order variation 
            and the gradient of the xc potential
            \begin{IEEEeqnarray}{rCl}
            V_\mathrm{xc}^{(1)\beta j+}(\mathbf{r})=n^{(1)\beta j+}(\mathbf{r})f_{\mathrm{xc}}(\mathbf{r}),~~~~\nabla_{j}V_\mathrm{xc}(\mathbf{r})=\left(\nabla_{j}n(\mathbf{r})\right)f_{\mathrm{xc}}(\mathbf{r}),
            \end{IEEEeqnarray}
            both the first-order variation of the charge density and the gradient of the unperturbed density are  multiplied with the exchange-correlation kernel 
            \begin{IEEEeqnarray}{rCl}
                f_{\mathrm{xc}}(\mathbf{r}) \coloneq \frac{\mathrm{\delta}V_{\mathrm{xc}}[n(\mathbf{r})]}{\mathrm{\delta}n(\mathbf{r})}|_{n^{(0)}(\mathbf{r})} \, ,
            \end{IEEEeqnarray}
            that is the functional derivative of xc-potential with respect to the charge density evaluated at the DFT ground state density $n^{(0)}(\mathbf{r})$ of the unperturbed system. Algorithmically, all operations are carried out in real space after the respective IR and MT coefficients of the density response and density gradients in the coefficient space (see definition~\eqref{eq:simplyrho} for clarity) have been transformed to real space by Fourier transformation and the evaluation of the lattice harmonics~\cite{Altmann:65} on a spherical grid, respectively. The results of the multiplication are then transformed back to coefficient space. 
            
            For the sake of algorithmic locality we recalculate the xc-kernal at each Sternheimer iteration again, although it depends only on the ground-state density and does not change with iteration. For the sake of convenience, we employ the \texttt{libxc} library of functionals~\cite{LehtolaSteigemannOliveiraEtAl2018} making all necessary quantities readily available when provided with the real-space density.
            Currently, we are limited to LDA functionals. In the future we plan also an extension to GGA functionals, for which the evaluation of the xc kernel is significantly more involved.
        \subsubsection{Achieving Self-Consistency} \label{sec:mixing}
            The self-consistent solution for the charge-density response field by means of  the Sternheimer equation bares a lot similarities to the self-consistent solution of the charge density by means of the Kohn--Sham equation in a conventional DFT calculation. In both cases we deal with a nonlinear problem that is solved iteratively. The output response density $n^{(1)}_{(m+1)}$, here and below written as scalar quantity as each atom and displacement coordinate is converged independently, obtained after completing iteration step $m$ is a functional of the first-order change of the potential $V^{(1)}_\mathrm{eff}$, which depends on the input response density $n^{(1)}_{(m)}$, which enters the Sternheimer equation to generate the wave function response from which the charge-density response  $n^{(1)}_{(m+1)}$ is calculated. Therefore, we adopted the existing charge-density mixing technology~\cite{Winkelmann:20} and mixed the charge-density response 
            \begin{IEEEeqnarray}{rCl}
            n^{(1)}_{(m+1)} = \textsc{Mix}\left(n^{(1)}_{(m)}, \delta n^{(1)}_{(m)},\alpha_{\mathrm{mix}}\right) 
            \end{IEEEeqnarray}
            according to the Broyden-like scheme of Anderson~\cite{mix_and} with the same mixing parameter $\alpha_{\mathrm{mix}}$ as in the ground-state calculation. $\delta n^{(1)}_{(m)}$ is the Anderson-preconditioned residual response density $ F\left(n^{(1)}_{(m)}\right)- n^{(1)}_{(m)}=n_{\mathrm{out}}^{(1)} - n_{\mathrm{in}}^{(1)}$ 
            with the preconditioner synthesized from the history of all charge density responses letting $n^{(1)}_{(m)}$ and $F\left(n^{(1)}_{(m)}\right)$ the input and output response charge density at some iteration $m$, respectively. To cope with the fact that in the MT region the density response is complex valued while  the original charge density is real valued, we mix the real and imaginary part of the response density independently by mapping both onto the mixing scheme for a magnetic system relating the real and imaginary part of the density response to spin-up and -down densities. In reality, the response density depends also on the ground-state charge densities. Throughout the self-consistency cycles, however, all ground-state properties remain unchanged and thus this charge-density field is taken off from the mixing procedure, as it provides a constant static offset that might contribute to an instability of the procedure. After the mixing is completed these terms are added again.  
            As a measure of convergence we use the $L_2$-norm induced metric
            \begin{IEEEeqnarray}{rCl}
              \mathrm{distance}\left(n_{\mathrm{out}}^{(1)\beta j+},n_{\mathrm{in}}^{(1)\beta j+}\right)=\left(\frac{1}{\mathrm{\Omega}}\int_{\mathrm{\Omega}}\left\|\,n_{\mathrm{out}}^{(1)\beta j+} - n_{\mathrm{in}}^{(1)\beta j+}\,\right\|\mathrm{d}\mathbf{r}\right)^{1/2}
              \label{eqn:distance}
            \end{IEEEeqnarray}
            and require it to be smaller than a preset threshold of $\epsilon_{\mathrm{scf}}$.
            All tests done to this point indicate very stable convergence behaviour for any material which converged properly in its ground-state calculation.
            But it should be noted that for the first few iterations the distance can start from very large values, especially when dealing with small $\mathbf{q}$-vectors. Some additional details on the mixing are found in \ref{sec:app_mix}.
   
    \subsection{Dynamical Matrix} \label{sec:dynmat}
    We recall, that according to~\eqref{eq:DM} and~\eqref{eq:FCM}, the DM is related to the second derivative of the Born-Oppenheimer energy surface. From~\eqref{eqn:verysimpleDM} we have seen, that this second derivative is related to the Coulomb interaction between the charge density response $n^{(1)}$ and the perturbed external potential $V_{\mathrm{ext}}^{(1)}$ generated by the nuclear charge, the Coulomb interaction interaction between the ground-state density $n$ and the second order external potential $V_{\mathrm{ext}}^{(2)}$, and the second derivative of the repulsive Coulomb energy generated by the nuclear charges. The Hellmann-Feynman force constant~\eqref{eqn:verysimpleDM} is an important contribution to the DM, but it is incomplete for many electronic structure methods, in particular for the LAPW basis set. In the following we derive the DM step by step starting with the first derivative of the Born-Oppenheimer energy surface. Of course, by this we find the Hellmann-Feynman terms again, but also the Pulay terms, and the terms due to the discontinuity at the MT-sphere boundary.  
    
    The ground-state energy~\eqref{eq:KS-energy} per unit cell of volume $\mathrm{\Omega}$ of the unperturbed system can equivalently be expressed in terms of the Kohn--Sham eigenvalues as
        \begin{IEEEeqnarray}{rCl}
            E_{\mathrm{tot}}
            &=&\sum_{\mathbf{k}\nu}\tilde{f}_{\mathbf{k}\nu}\epsilon_{\mathbf{k}\nu}-TS-\int_{\mathrm{\Omega}} n(\mathbf{r})V_{\mathrm{eff}}(\mathbf{r})\mathrm{d}\mathbf{r}+\int_{\mathrm{\Omega}} n(\mathbf{r})V_{\mathrm{ext}}(\mathbf{r}) \mathrm{d}\mathbf{r} \\
            &&{}+\frac{1}{2}\int_{\mathrm{\Omega}} n(\mathbf{r})V_{\mathrm{H}}(\mathbf{r})\mathrm{d}\mathbf{r}+\int_{\mathrm{\Omega}} n(\mathbf{r})\epsilon_{\mathrm{xc}}[n(\mathbf{r})]\mathrm{d}\mathbf{r}+\frac{1}{2}\sum_{\alpha\neq\beta}\frac{Z_{\alpha}Z_{\beta}}{|\boldsymbol{\tau}_{\alpha}-\boldsymbol{\tau}_{\beta}|} \,,\IEEEnonumber
        \end{IEEEeqnarray}
        where the first term together with the third one corresponds to the kinetic energy $T_0$, the fifth term is the Hartree energy $E_\mathrm{H}$, the sixth term the exchange-correlation energy $E_\mathrm{xc}$ and  the last term the Coulomb energy $E_\mathrm{ion-ion}$ between nuclei of atoms $\alpha$ and $\beta$ with atomic numbers $Z_{\alpha},Z_{\beta}$ of the energy functional~\eqref{eq:KS-energy}.  We introduced a term dependent on the temperature $T$ and electronic entropy $S$ as proposed by Weinert and Davenport~\cite{PhysRevB.45.13709} to deal consistently with the temperature dependent Fermi-Dirac distribution of the occupation of electron states in case of metals.
        
        From this we derive an optimized representation of the first-order total energy variation. All contributions related to the first-order occupation numbers cancel  between the sum of the single particle energies $\epsilon_{\mathbf{k}\nu}$ and entropy terms, and we find for a displacement of atom $\alpha$ along coordinate $i$
        \begin{IEEEeqnarray}{rCl}\hspace{-2cm}
            E_{\mathrm{tot}}^{(1)\alpha i-}
            &=&\int_{\mathrm{\Omega}} n(\mathbf{r})V_{\mathrm{ext}}^{(1)\alpha i-}(\mathbf{r})\mathrm{d}\mathbf{r}+E_{\mathrm{ion-ion}}^{(1)\alpha i-} \IEEEnonumber \\\
            &&{}+\sum_{\mathbf{k}\nu}\tilde{f}_{\mathbf{k}\nu}C_{\mathbf{k}\nu}^{(1)\alpha i-}+\int_{\mathrm{MT}^{\alpha}}n(\mathbf{r})\nabla_{\!\!i}V_{\mathrm{C}}(\mathbf{r})\mathrm{d}\mathbf{r} \IEEEyessubnumber\label{eq:E1full} \\
            &&{}+\int_{\mathrm{\Omega}}\Theta_{\mathrm{IR}}^{(1)\alpha i-}(\mathbf{r})n(\mathbf{r})\left( V_{\mathrm{C}}(\mathbf{r})+\epsilon_{\mathrm{xc}}[n(\mathbf{r})]\right)\mathrm{d}\mathbf{r} \, ,\IEEEnonumber \label{eq:E(1)}\\[2ex]
            \mathrm{with}\qquad\quad\quad C_{\mathbf{k}\nu}^{(1)\alpha i-}&\coloneq&\Braket{\widetilde{\Psi}_{\mathbf{k}\nu}^{i}|\mathscr{H}-\epsilon_{\mathbf{k}\nu}|\Psi_{\mathbf{k}\nu}
            }_{\!\!\!\!~\alpha}+\Braket{\Psi_{\mathbf{k}\nu} |\mathscr{H}-\epsilon_{\mathbf{k}\nu}|
            \widetilde{\Psi}_{\mathbf{k}\nu}^{i}}_{\!\!\!\!~\alpha}\qquad\qquad\qquad \IEEEyessubnumber \\
            &&{}+\Braket{\Psi_{\mathbf{k}\nu}^{\mathrm{IR}} |\Theta_{\mathrm{IR}}^{(1)\alpha i-}(\mathscr{T}-\epsilon_{\mathbf{k}\nu})|\Psi_{\mathbf{k}\nu}^{\mathrm{IR}}}_{\mathrm{\Omega}}. \IEEEnonumber
        \end{IEEEeqnarray} 
        $E_{\mathrm{tot}}^{(1)}$ is, aside from the explicit q-dependence in the first-order quantities and the fact that it is not evaluated for a finite displacement with amplitude $\mathbf{Q}$ (see \eqref{eq:DTau}), reminiscent of the LAPW force expression introduced by Yu and Krakauer~\cite{Yu1991_PhysRevB_43_6411} with the discontinuity extension of Klüppelberg \textit{et al.}~\cite{PhysRevB.91.035105}, and thus the $i$-th force component acting on atom $\alpha$, $F^\alpha_i$, is related to $E_{\mathrm{tot}}^{(1)\alpha i}(\mathbf{q}=\mathbf{0})$ as $F^\alpha_i=-E_{\mathrm{tot}}^{(1)\alpha i}(\mathbf{0}) \, Q_i(\mathbf{0})$. 
        The first line of \eqref{eq:E1full} corresponds to the well-known Hellmann--Feynman force and, like in the implemented form of the Sternheimer equation~\eqref{eqn:SternheimerEquationImplemented}, the Pulay and the MT surface-term contributions are smartly rearranged to discard gradients applied to \textit{e.g.}\ wave functions (by reformulation of the MT surface integrals into volume integrals of gradients). We arrive at (i) state-dependent correction terms $C_{\mathbf{k}\nu}^{(1)\alpha i-}$, which are a sum of typical Pulay-type and MT surface terms evaluated in the MT-sphere and IR, respectively, (ii) the potential energy of the ground-state charge density in the field of the gradient of the Coulomb potential $\boldsymbol{\nabla} V_\mathrm{C}$ in the displaced muffin-tin sphere, as well as (iii) the electrostatic energy of the charge density in the field of the Coulomb potential and the exchange correlation energy $E_\mathrm{xc}$ both evaluated in the IR with the perturbed step function. 
       
        Based on the same reformulation ideas, we obtain the following collection of terms for the second order change of the total energy per unit-cell volume with respect of the displacement of atom $\alpha$ into direction $i$ and atom $\beta$ into direction $j$:
        \begin{IEEEeqnarray}{rCl}
            E_{\mathrm{tot}}^{(2)\beta j+\alpha i-}
            &=&\int_{\mathrm{\Omega}} \left[n^{(1)\beta j+\nabla}(\mathbf{r})V_{\mathrm{ext}}^{(1)\alpha i-}(\mathbf{r})+(\nabla_{\!\!j}n(\mathbf{r}))V_{\mathrm{ext}}^{(1)\alpha i0\nabla}(\mathbf{r})\mathrm{\delta}_{\beta\alpha}\right]\mathrm{d}\mathbf{r} \IEEEnonumber\\
            &&{}-\int_{\mathrm{MT}^{\beta}} (\nabla_{\!\!j}n(\mathbf{r}))V_{\mathrm{ext}}^{(1)\alpha i-\nabla}(\mathbf{r})\mathrm{d}\mathbf{r}+\oint_{\partial \mathrm{MT}^{\beta}} n(\mathbf{r})V_{\mathrm{ext}}^{(1)\alpha i-}(\mathbf{r}) \uv{e}_{j}\mathrm{d}S \IEEEnonumber\\
            &&{}-\sum_{\gamma}\oint_{\partial \mathrm{MT}^{\gamma}} [n(\mathbf{r})V_{\mathrm{ext}}^{(1)\alpha i0}(\mathbf{r})]_{\mathrm{SF}}~ \uv{e}_{j}\mathrm{d}S\mathrm{\delta}_{\beta\alpha}\IEEEnonumber\\
            &&{}+\int_{\mathrm{\Omega}}\Theta_{\mathrm{IR}}^{(1)\beta j+}(\mathbf{r})n(\mathbf{r})V_{\mathrm{ext}}^{(1)\alpha i-}(\mathbf{r})\mathrm{d}\mathbf{r} + E_{\mathrm{ion-ion}}^{(2)\beta j+\alpha i-} \IEEEnonumber\\
            &&{}+\sum_{\mathbf{k}\nu}\left[\tilde{f}_{\mathbf{k}\nu}^{(1)\beta j+}C_{\mathbf{k}\nu}^{(1)\alpha i-}+\tilde{f}_{\mathbf{k}\nu}C_{\mathbf{k}\nu}^{(2)\beta j+\alpha i-}\right] \IEEEnonumber \IEEEnonumber \label{eqn:ccoeffs} \\
            &&{}+\int_{\mathrm{MT}^{\alpha}}\left[n^{(1)\beta j+\nabla}(\mathbf{r})\nabla_{\!\!i}V_{\mathrm{C}}(\mathbf{r})-(\nabla_{\!\!i}n(\mathbf{r}))V_{\mathrm{C}}^{(1)\beta j+\nabla}(\mathbf{r})\right]\mathrm{d}\mathbf{r} \IEEEnonumber \\
            &&{}+\oint_{\partial \mathrm{MT}^{\alpha}} n(\mathbf{r})V_{\mathrm{C}}^{(1)\beta j+\nabla}(\mathbf{r}) \uv{e}_{i}\mathrm{d}S \IEEEnonumber \\
            &&{}+\int_{\mathrm{\Omega}}\Theta_{\mathrm{IR}}^{(1)\alpha i-}(\mathbf{r})\left[ n^{(1)\beta j+}(\mathbf{r}) V_{\mathrm{eff}}(\mathbf{r})+n(\mathbf{r}) V_{\mathrm{C}}^{(1)\beta j+}(\mathbf{r})\right]\mathrm{d}\mathbf{r} \IEEEnonumber \\
            &&{}+\int_{\mathrm{\Omega}}\Theta_{\mathrm{IR}}^{(1)\alpha i0}(\mathbf{r})\left[(\nabla_{\!\!j}n(\mathbf{r})) V_{\mathrm{eff}}(\mathbf{r})+n(\mathbf{r}) (\nabla_{\!\!j}V_{\mathrm{C}}(\mathbf{r}))\right]\mathrm{\delta}_{\beta\alpha}\mathrm{d}\mathbf{r} \IEEEyesnumber \,.
        \end{IEEEeqnarray}
          This lengthy expression is the complete FLAPW analogue of the Hellmann-Feynman expression of the Hessian matrix~\eqref{eqn:verysimpleDM}. The first four lines constitute the Hellmann-Feynman part, where parts of the integral terms stem 
          from rearrangements by partial integration to avoid second order dyadic gradient terms ($\nabla_{j}\nabla_{i}$) and gradient terms of the perturbed quantities at the expense of additional surface integrals. This was done on account of the observation that such terms (resulting here from the double direct differentiation of the external potential) are very demanding for the radial integration and are a major source of numerical inaccuracies. The same rationale holds for the various integral terms in the bottom four lines. The $(\mathbf{k}\nu)$-dependent terms contain only the part of the IBC that is directly basis dependent and mixes Pulay and surface contributions. The composition of the coefficients $C_{\mathbf{k}\nu}^{(2)\beta j+\alpha i-}$ can be found in~\ref{sec:app_c1}.
        
        We derive the second-order variation of the ion-ion interaction, $E_{\mathrm{ion-ion}}^{(2)}$, following a scheme for the ground state energy already published by  Weinert~\cite{WeinertWimmerFreeman1982}, bearing similarities to the perturbed electronic potentials.
        Ultimately, we use 
        \begin{IEEEeqnarray}{rCl}
            E_{\mathrm{ion-ion}}^{(2)\beta j+\alpha i-} = 4 \mathrm{\pi} \sum_{\alpha} Z_{\alpha} &&\left[\sum_{{\mathbf{G}\, (\ne\mathbf{0})}} \frac{n_{\mathrm{ps}}^{\beta j+i-}(\mathbf{G})}{\left|\mathbf{G}\right|^2} \mathrm{j}_{0}(\left|\mathbf{G}\right| R_{\mathrm{MT}^\beta})\right. \\
            -\delta_{\alpha,\beta}&&\left.\sum_{{\mathbf{G}+\mathbf{q}\, (\ne\mathbf{0})}} \!\!\frac{n_{\mathrm{ps}}^{\beta j+i-}(\mathbf{G}+\mathbf{q})}{\left|\mathbf{G} + \mathbf{q}\right|^{2}} \mathrm{j}_{0}(\left|\mathbf{G} + \mathbf{q}\right| R_{\mathrm{MT}^\beta})\right] \IEEEnonumber ,
        \end{IEEEeqnarray}
        with 
        \begin{IEEEeqnarray}{rCl}
            n_{\mathrm{ps}}^{\beta j+i-}(\mathbf{G}+\mathbf{q}) &=& \frac{Z_{\beta}}{\mathrm{\Omega}} \left(2N + 7\right)!! \frac{\mathrm{j}_{N + 3}(\left|\mathbf{G} + \mathbf{q}\right| R_{\mathrm{MT}^\beta})}{\left(\left|\mathbf{G} + \mathbf{q}\right| R_{\mathrm{MT}^\beta}\right)^{N + 3}} \mathrm{e}^{-\mathrm{i} \left(\mathbf{G} + \mathbf{q}\right) \cdot \boldsymbol{\tau}_{\beta}} \left(G + q\right)_{j} \left(G + q\right)_{i} \IEEEyesnumber \,. \label{eqn:Eii2FourierCoeffcorr}\IEEEeqnarraynumspace
        \end{IEEEeqnarray}
        The parameter $N$ appearing in the pseudodensity, $n_{\mathrm{ps}}$, is chosen for its optimal convergence and that of the IR potential according to the given choice of $\ell_{\mathrm{max}}$ and $G_{\mathrm{max}}R_{\mathrm{MT,max}}$~\cite{Weinert1981_JMathPhys_22.2433}. We choose this parameter in the same way as for the calculation of the ground state. The expression for $n_{\mathrm{ps}}$ results from the evaluation of equation (28) in~\cite{Weinert1981_JMathPhys_22.2433} for $\ell=2$ with multipole coefficients representing the second order atomic displacements already calculated and expressed as factors containing the reciprocal wave vector components $(G+q)_{i/j}$. The power $N+3$ in the denominator is two orders higher than in the reference, which is compensated again by the factors $G+q$ occurring in the nominator resulting from the second-order differentiation of the energy. The numerical quality of this formalism is in good agreement with the results obtained from the \texttt{ABINIT}~\cite{GonzeBeukenCaracasEtAl2002, RomeroAllanAmadonEtAl2020, GonzeAmadonAntoniusEtAl2020} code, where the algorithm is based on an Ewald approach~\cite{GerhorstDr:}.
       
        We then set up the DM by symmetrizing the energy perturbation
        \begin{IEEEeqnarray}{rCl}
            \matr{E}_{\mathrm{tot, sym}}^{(2)+-}(\mathbf{q})=\frac{1}{2}\left(\matr{E}_{\mathrm{tot}}^{(2)+-}(\mathbf{q})+\matr{E}_{\mathrm{tot}}^{(2)+-\dagger}(\mathbf{q})\right)
        \end{IEEEeqnarray}
        to ensure its hermiticity
        and dividing each element by a factor of $\sqrt{M_{\alpha}M_{\beta}}$. Then, we calculate all eigenvalues $\{\lambda_{\mu}\}_{\mu=1,...,3N_\mathrm{A}}$ and eigenvectors of the Hermitian matrix using a standard eigenvalue solver~\cite{lapack99}.
        The actual phonon frequencies $\{\omega_{\mu}\}_{\mu=1,...,3N}$ are the square roots of these eigenvalues. 
        As described in the literature, if an eigenvalue is positive, we take the resulting square root with a positive sign, and if it is negative and thus the square root would yield an imaginary frequency, we represent it as real with a negative value in our calculated phonon dispersions. 
        At negative frequencies, the phonon spectrum thus indicates instabilities in the crystal lattice.
        A deeper insight into the technical nuances of the implementation (such as integral evaluations, pseudodensity coefficients, and gradient calulations) is provided
        as an integral part of references~\cite{KlueppelbergDr} and~\cite{GerhorstDr:}.

\subsection{Scaling behaviour}
The DFPT algorithm comes on top of a ground-state calculation whose computational effort was briefly discussed in section~\ref{sec:flapw} and for which a detailed discussion can be found in~\cite{10.1007/978-3-319-96983-1_52}.
The runtime determining step of the DFPT algorithm is the iterative solution of the Sternheimer equation~\eqref{eqn:SternheimerEquationImplemented}
for each wave vector $\mathbf{q}$, for all three Cartesian coordinates of the displacement perturbation, all $N_\mathrm{A}$ atoms in the unit cell, and all $N_{\mathbf{k}}$ k-points in the BZ.
In practice, this is done by a series of matrix multiplications and thus the computational effort is bounded by the largest among them. This is already the first one, where we multiply the perturbed Hamiltonian and overlap matrices ($N_{\mathrm{B}}\times N_{\mathrm{B}}$, where $N_{\mathrm{B}}$ is the number of basis functions as determined by $K_{\mathrm{max}}$) with the matrix of unperturbed expansion coefficients in the occupied subspace ($N_{\mathrm{B}}\times N_{\mathrm{o}}$), with the number of occcupied states $N_{\mathrm{o}}$. The order of operations for this multiplication is $\mathcal{O}(N_{\mathrm{o}}N_\mathrm{B}^2)$.
The other matrix multiplications are of the same order, as the dimension of the occupied subspace gets passed on with each product, and there is no proper matrix inversion necessary for the initial Hamiltonian and overlap, as we use the spectral representation for a quasi-analytic inversion. This is of the order $\mathcal{O}(N_{\mathrm{o}}N_\mathrm{B})$.
Summarizing, for each wave vector $\mathbf{q}$, the runtime of the DFPT algorithm scales as $\propto 3N_{\mathbf{k}} N_\mathrm{A}N_\mathrm{o}N_\mathrm{B}^2$.
Since the number of occupied states as well as the number of basis functions scale with the number of atoms, the DFPT has a volume scaling of $\mathcal{O}(N_{\mathrm{A}}^4)$ and the precision scaling is of $\mathcal{O}(N_\mathrm{B}^2)$ in the number of basis functions.
Although in the DFPT approach, the volume scaling is worse than for the conventional DFT self-consistency cycle ($\propto N_{\mathbf{k}} N_{\mathrm{B}}^3$), in the FLAPW method the number of occupied states are only a fraction of all $N_\mathrm{B}$, \textit{e.g.}\ in fcc Ne we find 4 occupied states for 162 to 177 states overall (depending on the k-point). This is at most 2.5\%. In general we expect a maximum occupancy in the order of 5-10\%. 
Currently we use all available unoccupied states ($N_{\mathrm{B}}-N_{\mathrm{o}}\simeq N_\mathrm{B}$) in calculating the response matrix. Thus, $N_{\mathrm{o}}$ produces a prefactor that is a fraction of $N_{B}$ and an iteration of the Sternheimer loop is faster than that of a conventional DFT calculation with no symmetry.
The memory requirement, as opposed to the ground-state calculation, is more than tripled. This is due to the necessity of not only keeping the occupied unperturbed eigenvalues $\epsilon_{\mathbf{k}\nu}$ and eigenvectors $\mathbf{z}_{\mathbf{k}\nu}$ in storage, but also the full set of unperturbed $\epsilon_{\mathbf{k}+\mathbf{q},\nu'}$ and $\mathbf{z}_{\mathbf{k}+\mathbf{q},\nu'}$, as well as the occupied perturbed quantities $\epsilon_{\mathbf{k}+\mathbf{q},\nu}^{(1)\beta j}$ and $\mathbf{z}_{\mathbf{k}+\mathbf{q},\nu}^{(1)\beta j}$. The q-dependent quantities, however, can be deleted once a specific q-point calculation is finished.
\section{Results and Discussion} \label{sec:results}
    In this section we validate our DFPT framework with respect to the quality of Goldstone modes and phonon dispersion relations against the FD approach. We choose a set of six distinct elemental materials, none of which share the same attributes. We cover a simple alkali metal (Na), several magnetic (Fe, Ni)  transition and noble metals (Cu) exhibiting different crystal structures, a semiconductor (Si) and an insulating noble gas crystal (Ne). The alkali metal is distinguished by its rather simple Fermi surface with a low number of electrons, and the noble gas crystal by its low-energy phonon modes.
Our strategy is the following: We set up the unperturbed unit cell of the material under study and optimize its volume (fit the total energy curve as a function of different lattice constants to the Birch--Murnaghan equation of states~\cite{PhysRev.71.809}) and internal degrees of freedom of the atom positions if necessary.
We use the resulting structure as input for comparative FD calculations with \texttt{phonopy} and DFPT in \texttt{FLEUR}. We will first give a short summary on how the FD calculations are conducted.

    \subsection{Computational Details: Finite Displacement Phonons}
    To begin a FD calculation, we provide the  unit cell optimized by \texttt{FLEUR} as input to \texttt{phonopy} along with a 3$\times$3 matrix of integers, $\matr{M}_{\mathrm{S}}$, that extends the original Bravais lattice,  $\matr{A}_{\mathrm{u}}$, to a supercell  with lattice vectors $\matr{A}_{\mathrm{S}}$,  by the matrix multiplication: $\matr{A}_{\mathrm{S}}=\matr{A}_{\mathrm{u}}\matr{M}_{\mathrm{S}}$.
    The supercell is subsequently filled with copies of the original set of atoms at appropriate positions.
    To ensure the best possible comparability between our benchmarks, we set a list of computational parameters identically for all materials considered (Table~\ref{tab:parameters}) and we work with the same k-point densities across all different Brillouin zones in use. We have chosen a k-point set of $16\times 16\times 16$ for the ground-state calculations performed in the primitive unit cell and adjusted the k-point set for the supercells accordingly. As default size of the supercell we chose $2\times 2\times 2$ times the primitive unit cell with a k-point set reduced to $8\times 8\times 8$. For the $4\times 4\times 4$ supercell we work with a $4\times 4\times 4$-k-point set.  
    Aside from parameters previously mentioned, there is  the force convergence criterion $\epsilon_{\mathrm{force}}$ (similar to $\epsilon_{\mathrm{scf}}$, but for the difference between the forces in two iterations), and the force level (0 means there will be no corrections as described in reference~\cite{PhysRevB.91.035105}, as we do not expect significant drift forces emerging for the selected materials). The same parameters are used for all calculations, \textit{i.e.}\ the ground-state calculation, the supercell ground-state and force calculation for the FD supercell, and the DFPT run for all systems discussed here.

\begin{table}
\caption{\label{tab:parameters}Overview of the shared calculational parameters for the selected materials. Parameters not contained in the table are kept at the \texttt{FLEUR} default or are explicitly mentioned when presenting the respective calculation results.}
\begin{tabular}{@{}llllll}
\br
$K_{\mathrm{max}}$ & $G_{\mathrm{max}}$ & $k_{\mathrm{B}}T$ & $\ell_{\mathrm{max}}$ & $\ell_{\mathrm{max,nsph}}$ & $N_{\mathrm{MT}}$ \\
\mr
4.5/$a_{0}$ & 15.0/$a_{0}$ & 0.005 htr & 9 & 7 & 981 \\
\br
$N_{k_{x/y/z}}$ & $\alpha_{\mathrm{mix}}$ & $\epsilon_{\mathrm{scf}}$ & xc-functional & $\epsilon_{\mathrm{force}}$ & Force level \\ 
\mr
16 & 0.05 & 0.00001/$a_{0}^{3}$ & VWN\cite{vwn} & 0.00001 htr/$a_{0}$ & 0 \\
\br
\end{tabular}
\end{table}
    Along with a perfect supercell, \texttt{phonopy} analyzes the symmetry of the system and gives a list of supercell inputs with displacements that include all information  needed to construct the force constant matrix and thereby the dynamical matrix. This list of inputs goes back to the \texttt{FLEUR} code, which calculates the corresponding forces upon each suggested displacement. The force calculations for the different displacements are fully independent of each other, so the process can be run in parallel. Based on the set of force and displacement information, \texttt{phonopy} delivers the force constant matrix 
    and the final output is a continuous phonon dispersion relation based on a Fourier transform of this matrix onto reciprocal space. We acknowledge that the FD method contains harmonic and anharmonic contributions to the phonon-dispersion. The anharmonic contribution depends on the magnitude of the displacement amplitude. To compare our results with the DFPT, which contains strictly only the harmonic terms, we have carefully monitored the role of the displacement amplitude. Finally, we use a displacement amplitude of $0.02$~$a_{0}$ for each of the structures.
    
     \subsection{Computational Details: Density Functional  Perturbation Theory Phonons}\label{ssec:CD-DFPT}
     From the optimized \texttt{FLEUR} input cell, the DFPT calculation is started with the same computational parameter set as for the FD benchmark. It is important to note that the cutoff $K_{\mathrm{max}}$, which limits the number of reciprocal lattice vectors $\mathbf{G}$ for every k-point according to $|\mathbf{k}+\mathbf{G}|\le K_{\mathrm{max}}$ is also applied to the q-shifted k-points $\mathbf{k}+\mathbf{G}+\mathbf{q}$, just as the cutoff $G_{\mathrm{max}}$ is applied to  the density and potential responses. Practical experience has shown that we numerically obtain the best results when the q-points for which the phonon properties are calculated are part of the k-point mesh. Thus the choice of the selected q-points impacts also the choice of the equidistant k-point mesh. We would also like to point out that the differentiation of a function expanded into an angular momentum representation with angular momentum index $\ell$, also generates contributions in the angular momentum components of index $\ell\pm 1$. To fully capture these components, we increase the maximum angular momentum of the LAPW basis set from $\ell_\mathrm{max}$ in a DFT calculation, which is typically an even number, to $\ell_\mathrm{max}+1$ in the DFPT calculation, which explains the odd values of $\ell_\mathrm{max}$ in Table~\ref{tab:parameters}. 
     Analogously we proceed for the response charge density and potential. Also here we increase the angular  expansion to $L_\mathrm{max}+1$. By cubic point group symmetry, these angular momentum components are not occupied for ground state calculations and thus these quantities are not altered by increasing the cutoff by $1$.
     
     We first run a standard ground-state DFT calculation (red box in Figure~\ref{fig:scf1}) and, after the density is converged, modify the \texttt{inp.xml} file so that all states are taken into account in the eigenvalue determination (\texttt{numbands="all"}), and add a path with all q-points we want to evaluate in the \texttt{juPhon} tag. The calculations were performed for the \texttt{FLEUR} version that can be found on the repository under the Git tag \texttt{juBranch\_before\_DFPT\_merge}. A comprehensive description of the full workflow can additionally be found under the tag \texttt{Phonon\_README\_for\_paper}. Starting the \texttt{FLEUR} calculation with this modified input will calculate dynamical matrices for each q-point provided.
     
    \subsection{Quality of the Goldstone Modes}
    For any crystal with $N_{\mathrm{A}}$ atoms in the unit cell, the phonon spectrum will have $3N_{\mathrm{A}}$ distinct branches, three acoustic and $3(N_{\mathrm{A}}-1)$ optical ones, some of which might be degenerate depending on the crystal symmetry.
    Especially near the $\mathrm{\Gamma}$-point, $\mathbf{q} = \mathbf{0}$, the acoustic branches are related to the speed of sound in a material by their slope. At the $\mathrm{\Gamma}$-point, \textit{i.e.}\ at the infinite-wavelength limit, the  phonon reduces to a rigid translation of the solid, which does not cost any energy and the lowest three frequencies are required to be exactly zero summarized by the acoustic sum rule. In a FD calculation, this corresponds to a vanishing net force summed over all atoms (drift force)~\cite{KlueppelbergDr}. In DFPT, with the analytical solution of the Sternheimer equation for monoatomic materials~\eqref{eq:an_sol}, one can show~\cite{GerhorstDr:} that the dynamical matrix itself must vanish for $\mathbf{q} = \mathbf{0}$, hence making the acoustic phonons gapless Goldstone modes~\cite{Leutwyler_1994}. This is not the case for polyatomic solids, where the acoustic branches have finite value and the matrix is not $0$ in every element.
    
    With respect to the numerical approach taken here, which results to finite accuracy in the evaluation of all equations, this zero condition required by physics is usually not perfectly realized in practice, and in the development of many phonon codes one has chosen to explicitly enforce it by subtracting either the drift force for an FD approach or a diagonal matrix with the three lowest eigenvalues for the DFPT implementations. 
    Thus, evaluating the eigenvalue spectrum for the acoustic modes at $\mathrm{\Gamma}$-point and in particular their deviation from zero, is a numerical check of the quality of the Goldstone modes and constitutes a very good test for the accuracy of our calculations and whether such corrections are warranted. Table~\ref{tab:goldstone} summarizes the Goldstone mode quantities for each of our test systems as well as the material specific parameters. Each of the supercell calculations was carried out with a $2\times 2\times 2$ supercell. For each of the systems \texttt{phonopy} suggests excactly one displacement pattern.

\begin{table}
\caption{\label{tab:goldstone}Overview of the lattice constants, $a$,  in $a_{0}$, MT radii, $R_\mathrm{MT}$, in $a_{0}$, and acoustic $\mathrm{\Gamma}$-point modes, $\omega$, in 1/cm for the selected materials. No frequency exceeds an absolute value of 1.0 in units of 1/cm (or 0.12398~meV | 0.029979~THz, respectively) providing an upper bound for the error of the Goldstone mode $\omega(0)=0$.}
\resizebox{\textwidth}{!}{
\begin{tabular}{@{}lllllll}
\br
&\hphantom{$-$}Na&\hphantom{$-$}Fe&\hphantom{$-$}Ni&\hphantom{$-$}Cu&\hphantom{$-$}Si&\hphantom{$-$}Ne\\ 
\mr
$a$&\hphantom{$-$}7.651&\hphantom{$-$}5.209&\hphantom{$-$}6.466&\hphantom{$-$}6.651&\hphantom{$-$}10.206&\hphantom{$-$}7.586\\
$R_\mathrm{MT}$&\hphantom{$-$}2.6&\hphantom{$-$}2.2&\hphantom{$-$}2.2&\hphantom{$-$}2.2&\hphantom{$-$}2.1&\hphantom{$-$}2.5\\
\mr
$\omega_{1,\mathrm{FD}}$&\hphantom{$-$}$1.41\times 10^{-1}$&$-9.05\times 10^{-2}$&\hphantom{$-$}$5.64\times 10^{-2}$&\hphantom{$-$}$1.96\times 10^{-6}$&\hphantom{$-$}$6.86\times 10^{-2}$&$-1.59\times 10^{-6}$\\
$\omega_{2,\mathrm{FD}}$&\hphantom{$-$}$1.41\times 10^{-1}$&$-9.05\times 10^{-2}$&\hphantom{$-$}$5.64\times 10^{-2}$&\hphantom{$-$}$4.33\times 10^{-6}$&\hphantom{$-$}$6.86\times 10^{-2}$&$-7.98\times 10^{-7}$\\ 
$\omega_{3,\mathrm{FD}}$&\hphantom{$-$}$1.41\times 10^{-1}$&$-9.05\times 10^{-2}$&\hphantom{$-$}$5.64\times 10^{-2}$&\hphantom{$-$}$5.88\times 10^{-6}$&\hphantom{$-$}$6.86\times 10^{-2}$&\hphantom{$-$}$1.48\times 10^{-6}$\\
\mr
$\omega_{1,\mathrm{DFPT}}$&$-2.49\times 10^{-1}$&\hphantom{$-$}$8.29\times 10^{-1}$&$-3.70\times 10^{-1}$&$-1.22\times 10^{-1}$&\hphantom{$-$}$3.31\times 10^{-1}$&$-1.01\times 10^{-1}$\\
$\omega_{2,\mathrm{DFPT}}$&$-2.46\times 10^{-1}$&\hphantom{$-$}$8.29\times 10^{-1}$&$-3.70\times 10^{-1}$&$-1.22\times 10^{-1}$&\hphantom{$-$}$3.34\times 10^{-1}$&$-1.01\times 10^{-1}$\\
$\omega_{3,\mathrm{DFPT}}$&$-2.32\times 10^{-1}$&\hphantom{$-$}$8.29\times 10^{-1}$&$-3.69\times 10^{-1}$&$-1.22\times 10^{-1}$&\hphantom{$-$}$3.37\times 10^{-1}$&$-1.01\times 10^{-1}$\\
\br
\end{tabular}}
\end{table}

    The frequencies are overall small, though in general a bit larger for the DFPT case. It can also be seen, that in the FD case the modes are closest to zero for the simplest materials, fcc Cu and fcc Ne. The other materials contain either local orbitals, a spin-polarization or more than one atom in the unit cell. Since the above Goldstone-mode requirement is well met for all systems, we have come to the decision not to correct our spectrum by applying the acoustic sum rule. Furthermore, the deviation from $0$ can be seen as a measure of accuracy for the overall frequencies. 
    
    We note that the convergence behaviour of the density response is directly linked to that of the DFT ground-state calculation. For fast converging materials, the $\mathrm{\Gamma}$-point calculation will converge with similar speed.
    The calculations for other high-symmetry points in the phonon BZ require  some more self-consistency iterations and start with higher initial distances~\eqref{eqn:distance}.
    The calculation of intermediary $\mathbf{q}$-vectors takes even longer, with the iteration count growing noticeably with decreasing magnitude of $\mathbf{q}$.
    Overall, the calculations tend to finish in at most 15 iterations.
    
    \subsection{Comparison of Phonon Dispersion Relations}
    Here we validate our implementation of the DFPT by comparing the phonon dispersion relations of the materials introduced above with results from FD calculations along high-symmetry lines of the BZ. The results of the FD are shown as red dashed lines and the DFPT data points as blue squares. Since we deal with monoatomic systems (with the exception of Si), we find 3 acoustic modes that are partly degenerate. Overall we find an excellent agreement between the DFPT and FD approach. For some systems  we find unsatisfactory convergence at certain q-points. For these cases we investigate the convergence of the dispersion relation with respect to the increase  of the supercell size for FD calculations. 
    In the following we first present the alkali metal Na and the noble metal Cu, both having one valence $s$ electron, then we turn to the magnetic transition metals Fe and  Ni, and finally we present the covalently bonded semiconductor Si and the van-der-Waals bonded insulating noble gas crystal Ne. Since our main emphasis is on the numerical validation of our results, we do not discuss the physics of the lattice dynamics of the individual systems, but rather try to cover different classes of materials with our examples.
    
    We begin with bcc Na and fcc Cu to test the implementation for simple, non-magnetic metals. We restrict our DFPT calculations to $\mathbf{q}$-vectors that mediate between the k(')-points of the set sampling the first Brillouin zone, \textit{i.e.}\ $\mathbf{k}+\mathbf{q}=\mathbf{k}'+\mathbf{G}$, where $\mathbf{G}$ is an arbitrary reciprocal lattice vector.
    q-points unrelated to the k-point grid show a more erratic convergence behaviour and generally lead to unfavourable results. This gives us 24 distinct data points to compare our \texttt{phonopy} curves to:
    \begin{figure}[h]
        \includegraphics[width=\textwidth]{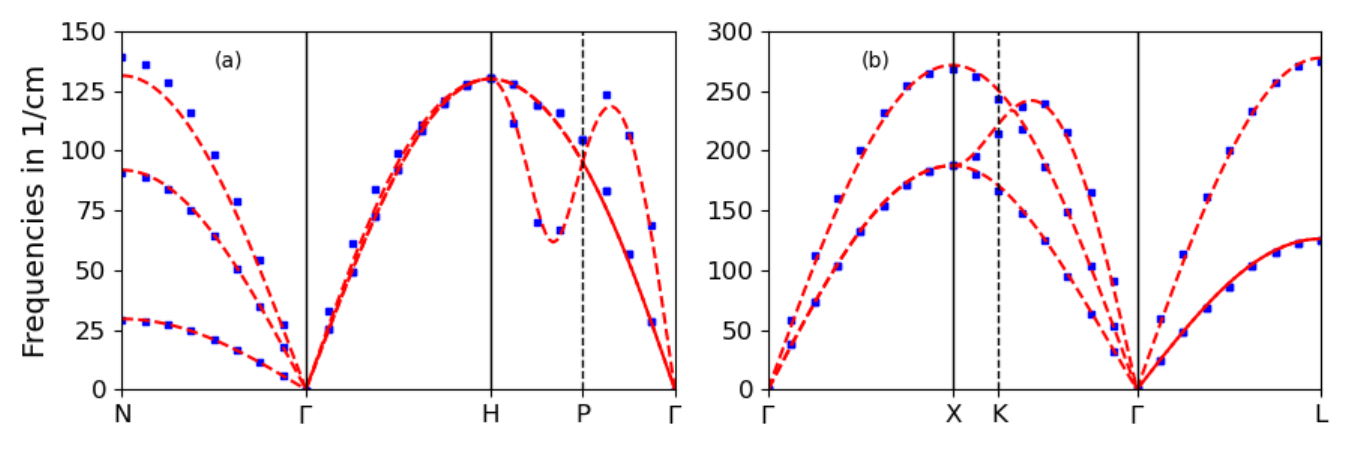}
        \caption{Phonon dispersions for (a) bcc Na and (b) fcc Cu. The red dashed curve shows the FD reference and the blue squares show the DFPT data. The MT-radii and lattice constants are (a) 2.6/7.651~$a_{0}$ and (b) 2.2/6.651~$a_{0}$. Both FD calculations are performed with a $2\times2\times2$ supercell. In the case of Na, the LAPW basis is supplemented with local orbitals (LO) to treat the 2s and 2p semicore states as valence states.}
        \label{fig:simple_metals}
    \end{figure}
    From Figure~\ref{fig:simple_metals} it can be seen that the overall agreement between FD and DFPT is good, but Cu matches more closely. In this context, it is useful to point out that the frequency scale of Na has twice the resolution of Cu. The Na DFPT data points deviate slightly from the FD curve for the upper (longitudinal) branch along the $\mathrm{N}$--$\mathrm{\Gamma}$-path, along the $\mathrm{\Gamma}$--$\mathrm{H}$ path the degeneracy between the longitudinal and transversal  branch is lifted, which can be recognized by two little blue squares at different frequency for each k-point, \textit{i.e.}\ a gap between both branches opens, which is a bit too big, and the high symmetry point $\mathrm{P}$ is not reproduced perfectly. A similar picture emerges for two ferromagnetic metals, fcc Ni and bcc Fe (Figure~\ref{fig:mag_metals}). 
     \begin{figure}[h!]
        \includegraphics[width=\textwidth]{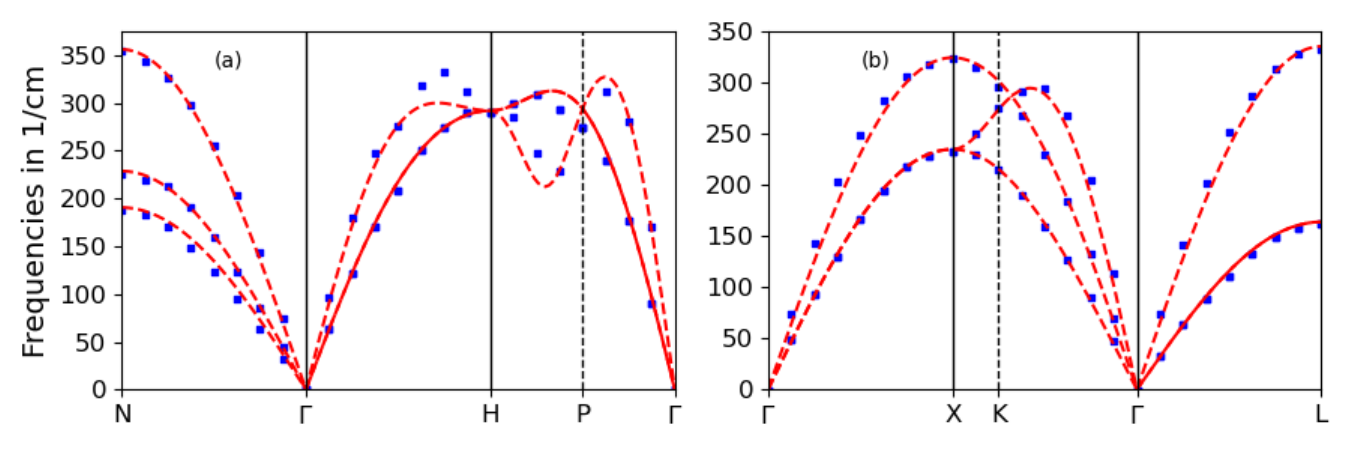}
        \caption{Phonon dispersions for ferromagnetic (a) bcc Fe and (b) fcc Ni. Colors and symbols like in Figure~\ref{fig:simple_metals}. The MT-radii and lattice constants are (a) 2.2/5.209~$a_{0}$ and (b) 2.2/6.466~$a_{0}$. Both FD calculations are performed with a $2\times2\times2$ supercell. In the case of Fe, the LAPW basis is supplemented with local orbitals (LO) to treat the 3s and 3p semicore states as valence states.}
        \label{fig:mag_metals}
    \end{figure}
    Once again, the agreement for the face-centered cubic material is better than for the body-centered one. Especially the peak left of the $\mathrm{H}$-point and the area right of it are not described well. We speculate that the discrepancy is caused by the FD curve. 
    
\begin{figure}
        \includegraphics[width=\textwidth]{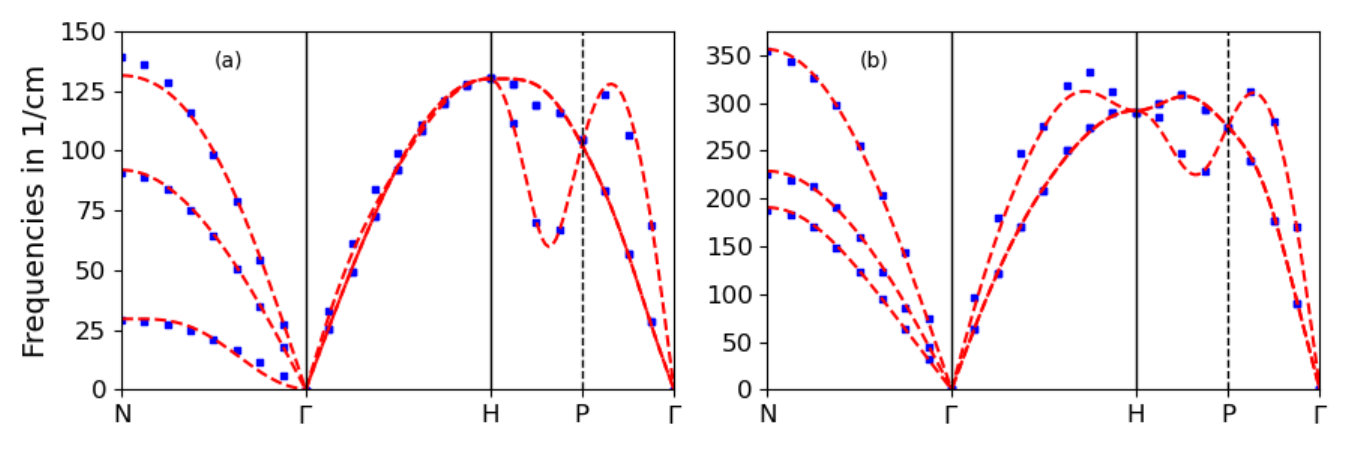}
        \caption{Improved phonon dispersions for (a) bcc Na and (b) bcc Fe. Colors and symbols like in Figure~\ref{fig:simple_metals}. Both FD calculations are performed with a $4\times4\times4$ supercell. The other parameters are left unchanged.}
        \label{fig:bcc_metals}
    \end{figure}

    To check for both bcc metals Na and Fe, whether the FD curves are not sufficiently converged in some regard and whether these discrepancies originate from the FD or DFPT part of the data, convergence tests were made. Differences, \textit{e.g.}\ between a $8\times8\times8$ and a $16\times16\times16$ k-point set were marginal. However, rerunning the FD calculations with a bigger $4\times4\times4$ supercell (this already constitutes a cell with 64 instead of 8 atoms)  leads to visibly improved results (Figure~\ref{fig:bcc_metals}).
    It is evident that the match between the curves and data points is neatly improved. We take away that certain materials may require larger supercells, but assume that they will converge slowly with respect to the supercell size and therefore omit further calculations here, as their compute time grows disproportionately.
    
     Finally, we show Si alongside fcc Ne, both FD calculations are carried out in the previous $2\times 2\times 2$-supercell, to have an example for a covalently bonded semiconductor and a van-der-Waals bonded insulator with low phonon frequencies. 
    \begin{figure}
        \includegraphics[width=\textwidth]{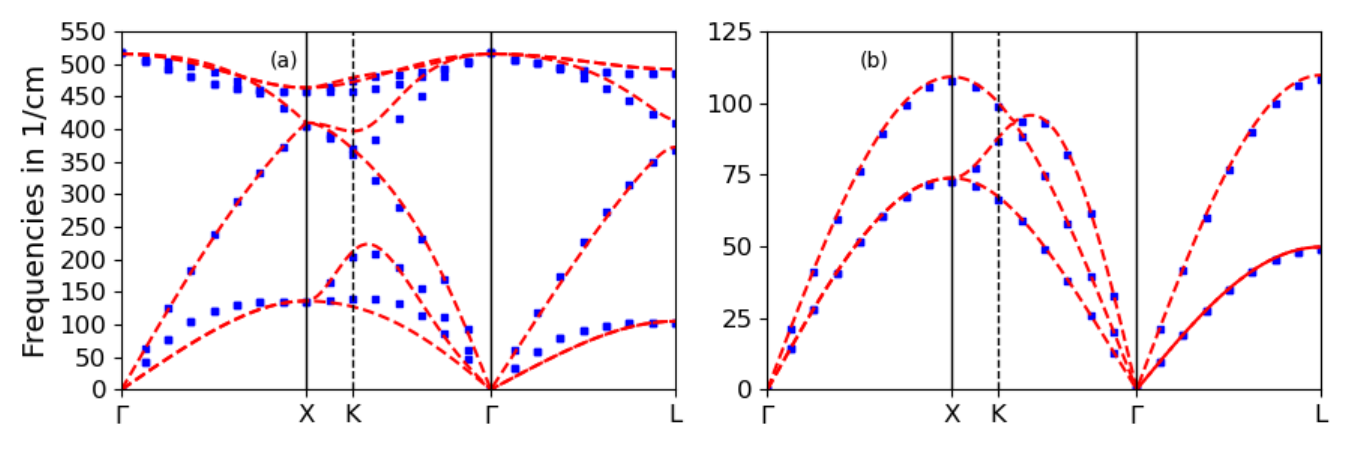}
        \caption{Phonon dispersions for (a) Si and (b) fcc Ne. Colors and symbols like in Figure~\ref{fig:simple_metals}. The MT-radii and lattice constants are (a) 2.1/10.206~$a_{0}$ and (b) 2.5/7.586~$a_{0}$. Both FD calculations are performed with a $2\times2\times2$ supercell.}
        \label{fig:iso_semi}
    \end{figure}
    For Si we came again  across the effect of an insufficient supercell size. The mismatch for certain Si branches is clearly visible, so to improve the fit we enlarge the supercell again. This time, we opt to use a more complex one, that reads $(M_{\mathrm{S}})_{ij}=2$ for $i\ne j$ and $(M_{\mathrm{S}})_{ii}=-2$ for $i\in\{1,2,3\}$. Again, we reduce the corresponding k-point set to $8\times8\times8$ points.
    It is equivalent to unfolding the diamond structure, fcc with a 2-atom basis, into a simple cubic supercell with 8 atoms and then duplicating it in each direction. This is computationally much cheaper than calculating a $4\times4\times4$ supercell, as the number of atoms in the unit cell is just half and the symmetry is reduced less by the single necessary perturbation in one atom. 
    \begin{figure}
        \includegraphics[width=\textwidth]{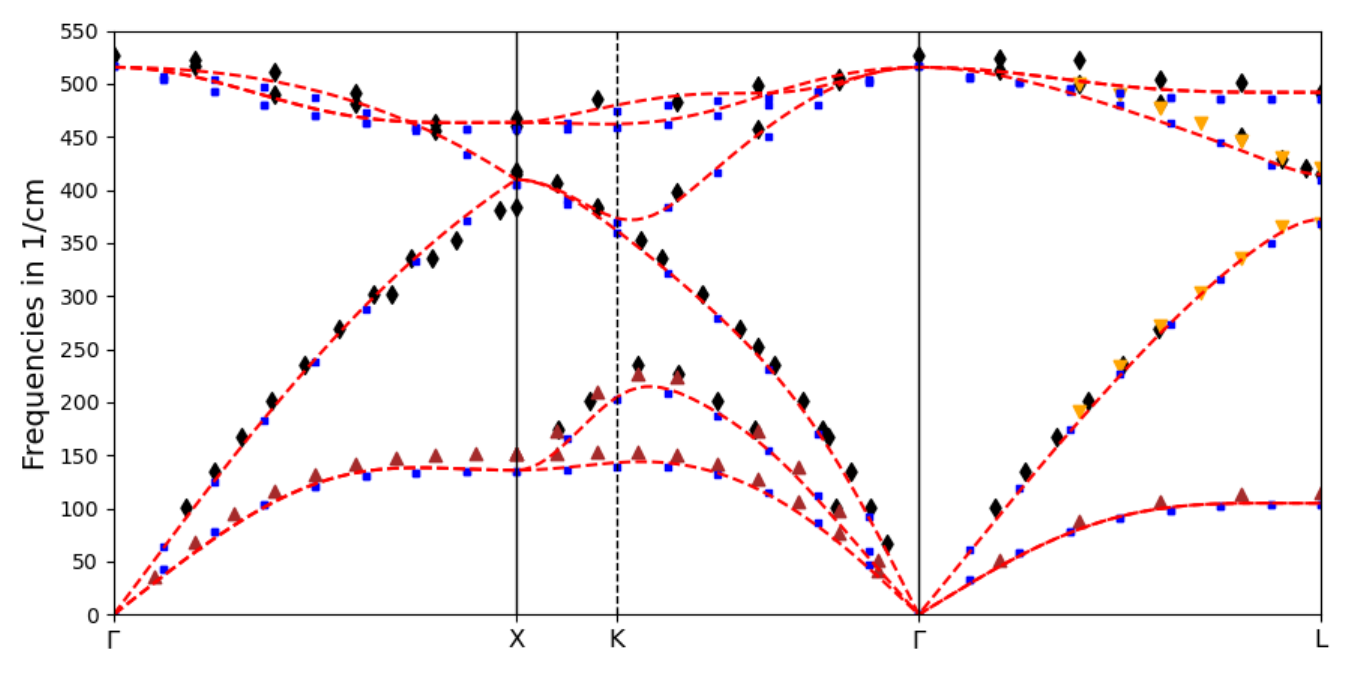}
        \caption{Improved phonon dispersion for Si with the modified supercell. The red curve shows the FD reference and the blue squares show the DFPT data. Additionally, we show three sets of experimental data. The black diamonds, brown upward triangles, and yellow downward triangles belong to references~\cite{PhysRevB.50.13347},~\cite{PhysRevB.6.3777}, and~\cite{Strauch1990} respectively.}
        \label{fig:silicon}
    \end{figure}
    The result is shown in Figure~\ref{fig:silicon}. It can easily be seen, that the larger supercell improves the overall match nicely. 
    The results of both methods give a good fit to the experimental data taken from various sources~\cite{PhysRevB.50.13347,PhysRevB.6.3777,Strauch1990} that we show together with our computational results, making both methods equivalently viable. A good agreement is obtained with reference calculations carried out with the normconserving pseudopotential method~\cite{Silicon_DFPT_ABINIT} in combination with  the LDA functional.
    
    Comparing the FD and DFPT results for fcc Ne, we find them in very good agreement, especially considering the small overall magnitude of the phonon dispersion. Although we focus in this paper on the internal consistency of the implementation of the DFPT, it is worth mentioning that for Ne the computational results do not agree well with the experimental data. Experiments at low temperatures~\cite{PhysRevB.11.1681} show a phonon dispersion that, when scaled to 1/cm, is roughly half as high in frequency at its maximum as the dispersion in Figure~\ref{fig:iso_semi}. This is understandable. Since Ne is a van-der-Waals bonded solid, we should have applied a van-der-Waals functional~\cite{Grimme-D3:10}. Using the conventional LDA, the Ne bonding becomes too strong and the phonon energy too high. This is consistent with the computationally  optimized lattice constant, which is around $3.924$~\AA, while the experimental data taken at $6$~K give a lattice constant of $4.466\pm0.002$~\AA. Just for comparison, the theoretical lattice constant of Si ($5.401$~\AA) matches the experimental one ($5.431$~\AA~\cite{articleSia}) quite well. To include the van-der-Waals functionals into the DFPT algorithm is part of our future plans.
\section{Conclusion and Outlook} \label{sec:conclusion}
     We presented an implementation of density-functional perturbation theory (DFPT) in the all-electron full-potential linearized augmented plane-wave (FLAPW) method \texttt{FLEUR} for the calculation of phonons, that is computationally stable and efficient. This complements the DFPT calculations of phonons, which are typically performed using pseudopotential methods with an all-electron approach, and extends an effective application of the  DFPT to magnetic systems and systems of localized electrons. The research software is built up modularly and can be extended in the future.  We developed and implemented algorithmic concepts to overcome or bypass numerical challenges inherent to the FLAPW concept, which are provided by the Madelung summation, the Coulomb singularity of the potential, the rapidly varying wave functions and charge densities in the vicinity of the nucleus, the calculations of gradients of the all-electron potential, the presence of the core electrons, the incompleteness and the position dependence of the basis-set, the different representations of the basis-set in muffin-tin-spheres and the interstitial region and their match at the muffin-sphere boundary to a point that the criterion for the Goldstone mode is satisfied to better than 0.125~meV. We highlighted some particularly challenging points and provided nitty-gritty details in how we dealt with them, leading to a collection of stable and accurate results validated by the finite difference (FD) method relying on accurate force calculations with respect to atomic displacements orchestrated by the phonopy software package~\cite{phonopy-phono3py-JPCM,phonopy-phono3py-JPSJ}. To achieve agreement between the FD and DFPT approaches, we noted the necessity of converging the FD calculations with respect to the supercell size, again confirming the quality of our DFPT results. 
    Considering the calculation of the phonon energy for the same three-dimensional grid of phonon wave vectors, at present, the FD approach shows a lower computational effort and takes less computer time than the DFPT method. This is also due to missing optimizations in the latter case, while for FD, the full symmetry of the atoms and forces can be exploited by the \texttt{FLEUR} code. It should be noted though, that the convergence of DFPT w.r.t.\ the k-point grid is much better than that of FD w.r.t.\ the supercell size. An in-depth optimization of the computational parameters with respect of the convergence of both methods in \texttt{FLEUR} will be conducted in the future.
    
    This paper serves as evidence that reliable and efficient phonon calculations with DFPT are possible in the FLAPW method. The computational efficiency can be further advanced by the full implementation of phonon symmetries~\cite{RevModPhys.40.1} as well as the implementation of effective parallelization strategies. The extension to polar materials~\cite{PhysRevB.1.910}, and the  implementation of the spin-orbit coupling~\cite{PhysRevB.78.045119}, non-collinear magnetism~\cite{PhysRevB.99.184404}, different exchange correlation functionals such as the generalized gradient approximation (GGA)~\cite{PhysRevLett.107.216402}, a van-der-Waals functional~\cite{Grimme-D3:10} or the extension to strongly correlated electrons systems using Hubbard $U$ (DFPT+U)~\cite{PhysRevB.101.064305} are straightforward and are subject to future work.
    
    Finally, the successful implementation of the DFPT formalism for phonons provides a motivation to translate the algorithm to other types of  perturbations, \textit{e.g.}\ to gain insight about responses to external electric~\cite{PhysRevB.75.115116} or magnetic fields~\cite{PhysRevB.99.184404}.
    
\section*{Acknowledgement} \label{sec:acknowledgement}
    We are grateful to Fabian Lux for his valuable contributions and insightful discussions. We gratefully acknowledge financial support by the European Centre of
    Excellence MaX ``Materials design at the Exascale'' (Grant No.\ 824143) funded by the EU, the Helmholtz Platform for Research Software Engineering - Preparatory Study (HIRSE\_PS), the Joint Lab Virtual Materials Design of the Forschungszentrum Jülich funded through the Innovation Fonds of the Federal Ministry of Education and Research (BMBF), the Joint Virtual Laboratory of the Forschungszentrum J\"ulich and the French Alternative Energies and Atomic Energy Commission — AI, Data Analytics and Scalable Simulation (AIDAS), and the Bavarian Ministry of Economic Affairs, Regional Development and Energy for financial support
within the High-Tech Agenda Project “Bausteine f\"ur das Quantencomputing auf Basis topologischer Materialien
mit experimentellen und theoretischen Ans\"atzen”. We  gratefully acknowledge computing time on the supercomputer JURECA~\cite{thornig2021jureca} at Forschungszentrum Jülich under grant no. jiff13. 

Finally, the authors dedicate this work to the memory of Henry Krakauer one of the original developer of the FLAPW method, teacher, advisor and mentor.

\section*{Data availability statement}
The data that support the findings of this study are available upon reasonable request from the authors and will be published on \texttt{zenodo.org} .

\appendix
\section{On the Symmetry of the Perturbed Wave Functions} \label{sec:app_psi1}
In~\eqref{eqn:rho1BasicTRS}, we presented the direct formulation of the density response. In practice, we instead make use of the time inversion symmetry $(\hat{T})$ and the space inversion symmetry $(\hat{P})$ of the $\mathbf{k}$-space:
    \begin{IEEEeqnarray}{rCl}
       \Psi_{\mathbf{k}\nu}^{*(1)\beta j+}(\mathbf{r})\Psi_{\mathbf{k}\nu}(\mathbf{r})&\overset{\hat{T}}{=}&\Psi_{-\mathbf{k}\nu}^{(1)\beta j+}(\mathbf{r})\Psi_{-\mathbf{k}\nu}^{*}(\mathbf{r})\overset{\hat{P}}{=}\Psi_{\mathbf{k}\nu}^{*}(\mathbf{r})\Psi_{\mathbf{k}\nu}^{(1)\beta j+}(\mathbf{r})\label{eq:app_TRS}\, ,
    \end{IEEEeqnarray}
    where we applied the component notation from~\eqref{eq:comp_not}. Then, we need only be concerned with quantities at $\mathbf{k}+\mathbf{q}$, not $\mathbf{k}-\mathbf{q}$, and  instead find
    \begin{IEEEeqnarray}{rCl}
            {n^{(1)\beta j+}_{\vphantom{{p}}}}(\mathbf{r}) &=& 2 \sum_{o} {\Psi_{\vphantom{f}o}^{*}}(\mathbf{r}) {\Psi_{\vphantom{f}o}^{(1)\beta j+}}(\mathbf{r}) \label{eqn:app_rho1BasicTRS} \,.
    \end{IEEEeqnarray}

    In situations where this symmetry is broken, such as in calculations involving spin-orbit coupling in combination with broken space-inversion symmetry, it may be necessary to perform a  calculation over the full Brillouin zone, \textit{i.e.}\ including the negative wave vectors, instead of relying on a prefactor of 2. This has already been implemented and utilized to confirm that Equation~\eqref{eq:app_TRS} holds, thus establishing a solid foundation for future implementations.

\section{Details on the First-Order Density Perturbation} \label{sec:app_rho1}
    In section~\ref{sec:rho1}, the calculation of terms depending on the perturbed occupation numbers $\tilde{f}_{\mathbf{k}\nu}^{(1)\beta j}$ are omitted. 
    They exclusively contribute to the $\mathrm{\Gamma}$-point phonons, \textit{i.e.}\ $\mathbf{q}=\mathbf{0}$, in case the unperturbed occupational numbers are fractional, and more than one atom is present in the unit cell (see~\ref{sec:app_occ} for further details).
    Provided the aforementioned conditions, the perturbed occupation-number terms are trivial in the sense that they couple to the unperturbed basis and produce charge density contribution to the response density.
    In the IR, we consequently add
    \begin{IEEEeqnarray}{rCl}
        n_{\mathrm{\mathrm{IR}},\tilde{f}^{(1)}}^{(1)\beta j}(\mathbf{r}) 
        \hspace{-0cm}&=& \frac{1}{\mathrm{\Omega}} \sum_{\mathbf{k}\nu} \tilde{f}_{\mathbf{k}\nu}^{(1)\beta j} \sum_{\mathbf{G}''} \sum_{\mathbf{G}'} z_{\mathbf{k}+\mathbf{G}',\nu}^{*} z_{\mathbf{k}+\mathbf{G}'',\nu} \mathrm{e}^{\mathrm{i} \left(\mathbf{G}'' - \mathbf{G}' \right) \cdot \mathbf{r}}. \IEEEyesnumber
    \end{IEEEeqnarray}
    In the MT spheres, the contributions can directly be absorbed into the $d$-coefficients that were referenced before.
    So, their full form is
    \begin{IEEEeqnarray}{rCl}
        d_{\ell' p' \ell'' p''}^{\gamma,\beta j}(L)&=&\sum_{m' m''} G^{m'', m, m'}_{\ell'', \ell, \ell'} \IEEEnonumber\\
        &&{}\times\left\lbrace\sum_{\mathbf{k}\nu} 2\tilde{f}_{\mathbf{k}\nu}
        A_{L' p'}^{\mathbf{k}\nu\gamma *} \bar{A}_{L'' p''}^{\mathbf{k}+\mathbf{q},\nu\gamma}(\beta)+\tilde{f}_{\mathbf{k}\nu}^{(1)\beta j}
        A_{L' p'}^{\mathbf{k}\nu\gamma *} A_{L'' p''}^{\mathbf{k}\nu\gamma}\right\rbrace \,, \label{eqn:mtRho1dCoeffs}
    \end{IEEEeqnarray}
    where the band-dependent matching coefficients enter as
    \begin{IEEEeqnarray}{rCl}
        A_{L' p'}^{\mathbf{k}\nu\gamma}&=&\sum_{\mathbf{G}}z_{\mathbf{k}+\mathbf{G},\nu}a_{L' p'}^{\mathbf{k}+\mathbf{G},\gamma},\IEEEyesnumber\IEEEyessubnumber\\
        \bar{A}_{L'' p''}^{\mathbf{k}+\mathbf{q},\nu\gamma}(\beta)&=&\sum_{\mathbf{G}}z_{\mathbf{k}+\mathbf{G}+\mathbf{q},\nu}^{(1)\beta j}a_{L'' p''}^{\mathbf{k}+\mathbf{G}+\mathbf{q},\gamma}+\delta_{\gamma\beta}\mathrm{i} (\mathbf{k} + \mathbf{G})z_{\mathbf{k}+\mathbf{G},\nu}a_{L'' p''}^{\mathbf{k}+\mathbf{G},\gamma}\IEEEyessubnumber \,. \label{eqn:mtRho1Coefficients}
    \end{IEEEeqnarray}

\section{Evaluating the Kinetic Energy Operator} \label{sec:app_ekin}
    There are several different ways of applying the kinetic energy operator $\mathscr{T}$ in an APW context. In deriving the Kohn--Sham equations, the variational expression of the kinetic energy of state $\nu$ reads 
    \begin{IEEEeqnarray}{rCl}
    T[\Psi_{\nu}]=\frac{1}{2}\int_{\mathrm{\Omega}} \boldsymbol{\nabla}\Psi^*_{\nu}(\mathbf{r})\cdot \boldsymbol{\nabla}\Psi_{\nu}(\mathbf{r})\,\mathrm{d}\mathbf{r}\,,
    \end{IEEEeqnarray}
    with first derivatives acting on the Kohn--Sham orbital $\Psi_{\nu}$ of state $\nu$. Conceptually and numerically, it is very convenient to determine the radial basis functions in the MT region as solutions of the Schrödinger equation. Therefore, by applying Green's theorem, one converts the representation of the kinetic energy in terms of the scalar product of two gradient terms into the Schrödinger form with the well-known Laplace operator, as in~\eqref{eqn:KSeqA}, 
    \begin{IEEEeqnarray}{rCl}
    T[\Psi_{\nu}]&&=\frac{1}{2}\int_\mathrm{IR} \boldsymbol{\nabla}\Psi^*_{\nu}(\mathbf{r})\cdot \boldsymbol{\nabla}\Psi_{\nu}(\mathbf{r})\,\mathrm{d}\mathbf{r}+\frac{1}{2}\int_\mathrm{MT} \Psi^*_{\nu}(\mathbf{r})(-\Delta) \Psi_{\nu}(\mathbf{r})\,\mathrm{d}\mathbf{r}\\
    &&+\frac{1}{2}\oint_{\partial \mathrm{MT}} \Psi^*_{\nu}(\mathbf{r})\boldsymbol{\nabla}\Psi_{\nu}(\mathbf{r})\,\cdot \mathrm{d}\mathbf{S}\,.\IEEEyesnumber\label{eq:lap_mix}
    \end{IEEEeqnarray}
    The latter term is the integral over the boundary $\partial \mathrm{MT}$ of each MT sphere, with the surface element $\mathrm{d}\mathbf{S}$ pointing outwards of the enclosed domain. Obviously, the surface term is zero if the wave function or its derivative is zero at domain boundary. This is in general not the case if the domain boundary is the surface between the MT and IR region. Applying the expression of the kinetic energy for the MT and IR region, we get the representation of the kinetic energy by the Laplace operator over the entire unit cell plus the difference of the surface terms at the muffin-tin spheres taken once from the domain of the MT and once from the domain of the IR region (for the definition of $\oint_{\partial \mathrm{MT}} [\mathbf{X}]\mathrm{d}\mathbf{S}$ see~\eqref{eqn:basicsurface})
    \begin{IEEEeqnarray}{rCl}
    T[\Psi_{\nu}]&&=\frac{1}{2}\int_{\mathrm{\Omega}} \Psi^*_{\nu}(\mathbf{r})(-\Delta) \Psi_{\nu}(\mathbf{r})\,\mathrm{d}\mathbf{r}+\frac{1}{2}\oint_{\partial \mathrm{MT}} [\Psi^*_{\nu}(\mathbf{r})\boldsymbol{\nabla}\Psi_{\nu}(\mathbf{r})]_{\mathrm{SF}}\,\cdot \mathrm{d}\mathbf{S}\,.\IEEEyesnumber\label{eq:lap_schrod}
    \end{IEEEeqnarray}
    In the limit of increasingly higher angular momentum $\ell_\mathrm{max}$ of the radial basis set in the muffin-tin sphere, the difference of the surface intergals converges to zero. In practice, we use finite $\ell_\mathrm{max}$ cutoffs and the surface integrals at the boundary discontinuity are finite and not negligible. 
    In the \texttt{FLEUR} code, we go one step further and symmetrize the form  \eqref{eq:lap_schrod} by applying the Laplace operator to both $\Psi_{\nu}$ and $\Psi_{\nu}^*$.
    \begin{IEEEeqnarray}{rCl}
    T[\Psi_{\nu}]&&=\frac{1}{4}\int_{\mathrm{\Omega}} \Psi^*_{\nu}(\mathbf{r})(-\Delta) \Psi_{\nu}(\mathbf{r})+\Psi_{\nu}(\mathbf{r})(-\Delta) \Psi^*_{\nu}(\mathbf{r})\,\mathrm{d}\mathbf{r}+T_{\mathrm{SF,sym}}[\Psi_{\nu}]\,.
    \label{eq:lap_schrod_fin}
    \end{IEEEeqnarray}
    The remaining symmetrized average surface contribution $T_{\mathrm{SF,sym}}$ is then negligible. It was tested for the DFPT implementation that there is no significant difference for calculations with the mixed form~\eqref{eq:lap_mix} as opposed to the symmetrized form~\eqref{eq:lap_schrod_fin}. We opt to use the latter for conformity with the base calculation.
    
\section{Modifying the Perturbed Expansion Coefficients} \label{sec:app_de}
One may naively think to ignore expression~\eqref{eqn:Sternheimershort} in case of tiny energy differences $\delta_{\mathbf{q}\nu',\mathbf{k}\nu}$. However, this is theoretically not correct and can cause numerical trouble at particular $\mathbf{q}$ vectors. Instead, in order to derive numerically stable forms  of $z_{\mathbf{q}\nu',\mathbf{k}\nu}^{(1)\beta j}$ we exploit the $\hat{T}$ and $\hat{P}$ symmetry between pairs of occupied states that enter the sum. We inspect the respective part of the first order density response:
\begin{IEEEeqnarray}{rCl}
                n^{(1)\beta j+}(\mathbf{r})
                &=&  \sum_{\mathbf{k}\nu} \tilde{f}_{\mathbf{k}\nu} \Psi_{\mathbf{k}\nu}^{*}(\mathbf{r})\sum_{\mathbf{G}}z_{\mathbf{k}+\mathbf{G}+\mathbf{q},\nu}^{(1)\beta j}\phi_{\mathbf{k}+\mathbf{G}+\mathbf{q}}(\mathbf{r})\IEEEnonumber \\
                &=&\sum_{\mathbf{k}\nu} \tilde{f}_{\mathbf{k}\nu} \Psi_{\mathbf{k}\nu}^{*}(\mathbf{r})\sum_{\nu'}\Psi_{\mathbf{k}+\mathbf{q},\nu'}(\mathbf{r})z_{\mathbf{q}\nu',\mathbf{k}\nu}^{(1)\beta j}\label{eqn:app_rho1short}\\
                &=&n_{\mathrm{occ}-\mathrm{occ}}^{(1)\beta j+}(\mathbf{r})+n_{\mathrm{occ}-\mathrm{unocc}}^{(1)\beta j+}(\mathbf{r})\,.\IEEEnonumber
            \end{IEEEeqnarray}
We take a closer look at the occupied--occupied subspace and introduce a factor $1=1-F(\epsilon_{\mathbf{k}+\mathbf{q},\nu'})+F(\epsilon_{\mathbf{k}+\mathbf{q},\nu'})$ to find
\begin{IEEEeqnarray}{rCl}
    n_{\mathrm{occ}-\mathrm{occ}}^{(1)\beta j+}(\mathbf{r})=&&\sum_{\mathbf{k}\nu} f_{\mathbf{k}}F(\epsilon_{\mathbf{k}\nu}) \Psi_{\mathbf{k}\nu}^{*}(\mathbf{r})\IEEEnonumber\\&&
    {}\times \sum_{\nu'|\mathrm{occ}}(1-F(\epsilon_{\mathbf{k}+\mathbf{q},\nu'})+F(\epsilon_{\mathbf{k}+\mathbf{q},\nu'}))\Psi_{\mathbf{k}+\mathbf{q},\nu'}(\mathbf{r})z_{\mathbf{q}\nu',\mathbf{k}\nu}^{(1)\beta j}\label{eqn:app_rho1occocc}\,.
\end{IEEEeqnarray}
While the part of the equation with $1-F(\epsilon_{\mathbf{k}+\mathbf{q},\nu'})$ is computed just like before (see~\eqref{eqn:Sternheimershort}), the remainder features a useful antisymmetric relation:
\begin{IEEEeqnarray}{rCl}
    &&\sum_{\mathbf{k}\nu,\nu'|\mathrm{occ}}f_{\mathbf{k}}F(\epsilon_{\mathbf{k}\nu}) \Psi_{\mathbf{k}\nu}^{*}F(\epsilon_{\mathbf{k}+\mathbf{q},\nu'})\Psi_{\mathbf{k}+\mathbf{q},\nu'}(-1)\frac{H_{\mathbf{q}\nu',\mathbf{k}\nu}^{(1)\beta j+}-\epsilon_{\mathbf{k}\nu}S_{\mathbf{q}\nu',\mathbf{k}\nu}^{(1)\beta j+}}{\epsilon_{\mathbf{k} + \mathbf{q},\nu'} - \epsilon_{\mathbf{k}\nu}}\\
    =-&&\sum_{\mathbf{k}\nu,\nu'|\mathrm{occ}}f_{\mathbf{k}}F(\epsilon_{\mathbf{k}\nu}) \Psi_{\mathbf{k}\nu}^{*}F(\epsilon_{\mathbf{k}+\mathbf{q},\nu'})\Psi_{\mathbf{k}+\mathbf{q},\nu'}(-1)\frac{H_{\mathbf{q}\nu',\mathbf{k}\nu}^{(1)\beta j+}-\epsilon_{\mathbf{k}+\mathbf{q},\nu'}S_{\mathbf{q}\nu',\mathbf{k}\nu}^{(1)\beta j+}}{\epsilon_{\mathbf{k} + \mathbf{q},\nu'} - \epsilon_{\mathbf{k}\nu}}\,.
\end{IEEEeqnarray}
If we define the left hand side as $a$ and the right hand side as $b$, we can use $a=(a+b)/2$, and write:
\begin{IEEEeqnarray}{rCl}
    a=\sum_{\mathbf{k}\nu,\nu'|\mathrm{occ}}f_{\mathbf{k}}F(\epsilon_{\mathbf{k}\nu}) \Psi_{\mathbf{k}\nu}^{*}F(\epsilon_{\mathbf{k}+\mathbf{q},\nu'})\Psi_{\mathbf{k}+\mathbf{q},\nu'}(-1)S_{\mathbf{q}\nu',\mathbf{k}\nu}^{(1)\beta j+}\,.
\end{IEEEeqnarray}
This directly corresponds to evaluating~\eqref{eqn:SternheimerEquationModded2} for the expansion coefficients. A similar train of thought (without the inserted factor) can be followed for $\delta_{\mathbf{q}\nu',\mathbf{k}\nu}\approx0$ by recognizing that the occupation prefactor will then be the same for both the original and the shifted Bloch vector $\mathbf{k}$. This leads to
\begin{IEEEeqnarray}{rCl}
    a_{\delta_{\mathbf{q}\nu',\mathbf{k}\nu}\approx0}=\sum_{\mathbf{k}\nu,\nu'|\mathrm{occ}}\tilde{f}_{\mathbf{k}}\Psi_{\mathbf{k}\nu}^{*}\Psi_{\mathbf{k}+\mathbf{q},\nu'}(-1)S_{\mathbf{q}\nu',\mathbf{k}\nu}^{(1)\beta j+} \,,
\end{IEEEeqnarray}
and consequently to~\eqref{eqn:SternheimerEquationModded}, removing the problem of divergent reciprocal energy terms.

\section{Details of the generation of the Coulomb potential response and gradient}
\label{sec:app_pot}
As mentioned in section~\ref{sec:v1}, the calculation of the Coulomb potential response, $V^{(1)}_\mathrm{C}$, and the Coulomb potential gradient, $\boldsymbol{\nabla} V_\mathrm{C}$, is largely analogous to the description in Ref.~\cite{Weinert1981_JMathPhys_22.2433}, when the density is replaced by the density response or the charge-density gradient, respectively. This appendix serves to outline the differences to the conventional generation of the Coulomb potential in a ground state calculation.  

Firstly, there are the surface corrections to the multipole moments $q_{\ell m}^{ \gamma\mathbf{R},\mathrm{SF}}$ in the MT sphere and in the IR, $ q_{\ell m}^{ \gamma\mathbf{R}I,\mathrm{SF}}$. 
Concerning the Coulomb potential response, $V_\mathrm{C}^{(1)\beta j +}$, they result from the displacement of atom $\beta$ into the direction  $j$  by a phonon with wave vector $\mathbf{q}$. For an atom at $\boldsymbol{\tau}_{\gamma}$ in unit cell $\mathbf{R}$, the MT contribution reads
\begin{IEEEeqnarray}{rCl}
       q_{\ell m}^{\beta j +| \gamma\mathbf{R},\mathrm{SF}}&\coloneq&\mathrm{\delta}_{\gamma \beta} \mathrm{e}^{\mathrm{i} \mathbf{q} \cdot \mathbf{R}} R_{\mathrm{MT}^{\gamma}}^{\ell + 2} \sum_{\ell'm'} \left[ n\right]_{\mathrm{MT}^{\gamma},\ell'm'}(R_{\mathrm{MT}^{\gamma}}) \sum_{m'' = -1}^{1}\zeta_{j,m''} G_{\ell,\ell',1}^{m, m', m''},\IEEEyesnumber\IEEEyessubnumber\label{eq:app_qSF1_MT}
\end{IEEEeqnarray}
where we omit here explicitly the transformation of the density representation in the sphere  from lattice harmonics denoted as $[\,\cdot\,]$, to spherical harmonics, and the interstitial contribution reads
\begin{IEEEeqnarray}{rCl}
  q_{\ell m}^{\beta j +| \gamma\mathbf{R} I,\mathrm{SF}}&\coloneq& \mathrm{\delta}_{\gamma\beta} \mathrm{e}^{\mathrm{i} \mathbf{q} \cdot \mathbf{R}} \sum_{\ell'm'} 4 \mathrm{\pi} \mathrm{i}^{\ell'} \sum_{\mathbf{G}} \mathrm{e}^{\mathrm{i} \mathbf{G} \cdot \boldsymbol{\tau}_{\gamma}} n^{\mathrm{IR}}(\mathbf{G})  \IEEEyessubnumber\label{eqn:qL_SF} \\
  &\times& \mathrm{Y}_{\ell'm'}^{*}(\mathbf{\hat{G}}) \mathrm{j}_{\ell'}(\left|\mathbf{G}\right|R_{\mathrm{MT}^{\gamma}})\sum_{m'' = -1}^{1} \zeta_{j,m''} G_{\ell,\ell',1}^{m, m', m''}. \IEEEnonumber
\end{IEEEeqnarray}
In both cases we need Gaunt coefficients, $G_{\ell,\ell',1}^{m, m', m''}=\oint\mathrm{Y}_{\ell m}^{*}(\mathrm{\Omega})\mathrm{Y}_{\ell' m'}(\mathrm{\Omega})\mathrm{Y}_{1 m''}(\mathrm{\Omega})\mathrm{d}S$,  and a matrix $\matr{\zeta}$, that links the natural spherical tensorial  coordinates of the magnetic quantum number $m''$ with indices $\{-1,0,1\}$ to the Cartesian ones,
\begin{IEEEeqnarray}{rCl}
  \matr{\zeta} = \sqrt{\frac{2 \mathrm{\pi}}{3}}
  \begin{pmatrix}
    1  & 0  & - 1 \\
    \mathrm{i} & 0 & \mathrm{i}   \\
    0  & \sqrt{2} & 0
  \end{pmatrix}\,.
\end{IEEEeqnarray}

The structure factor $\mathrm{e}^{\mathrm{i} \mathbf{G} \cdot \boldsymbol{\tau}_{\gamma}}$ in~\eqref{eqn:qL_SF} results in the expression for the pseudo-density and Coulomb potential being evaluated with a reciprocal vector $\mathbf{G}+\mathbf{q}$ instead of $\mathbf{G}$. The same holds true for the multipole moments of the density response, which is the second main difference to the ground-state procedure. 

The nuclear charge contribution $Z_\gamma$ of atom $\gamma$ to the multipole moments reads
\begin{IEEEeqnarray}{rCl}
       q_{1 m}^{\beta j +| \gamma\mathbf{R},\mathrm{ext}}&\coloneq&-\mathrm{\delta}_{\gamma \beta} \mathrm{e}^{\mathrm{i} \mathbf{q} \cdot \mathbf{R}}\frac{3}{4\mathrm{\pi}}Z_{\gamma}\zeta_{j,m}.\IEEEyesnumber
\end{IEEEeqnarray}
It replaces the standard contribution to $q_{00}$ from the spherical Coulomb potential $Z_\gamma/r$ of the positively charged nuclei. Aside from these deviations, the procedure from the seminal paper~\cite{Weinert1981_JMathPhys_22.2433} can be followed. 

In the case of the Coulomb potential gradient, $\nabla_{j}V_\mathrm{C}$, no additional vector $\mathbf{q}$ appears and in comparison to~\eqref{eq:app_qSF1_MT},  the structure factor vanishes, there is no restriction $\delta_{\gamma\beta}$ to the displaced MT sphere $\beta$, and the expression changes sign. We find then  for the MT part of the surface correction:
\begin{IEEEeqnarray}{rCl}
       q_{\ell m}^{j| \gamma,\mathrm{SF}}&\coloneq&-R_{\mathrm{MT}^{\gamma}}^{\ell + 2} \sum_{\ell'm'} \left[ n\right]_{\mathrm{MT}^{\gamma},\ell'm'}(R_{\mathrm{MT}^{\gamma}}) \sum_{m'' = -1}^{1}\zeta_{j,m''} G_{\ell,\ell',1}^{m, m', m''}.\IEEEyesnumber
\end{IEEEeqnarray}
The changes to the IR part and to the nuclear term are analogous.

\section{Peculiarities of the Sternheimer Mixing} 
\label{sec:app_mix}
    Here are two technical notes about the mixing of the density perturbation during the Sternheimer self-consistency loop: 
    Firstly, we decided to mix only the density response  without the gradient part of the density that appears in the displaced MT spheres, as we then deal with a more well-behaved quantity, and the gradient does not change between iterations anyway. 
    Secondly, before the mixing starts, two initial cycles of the Sternheimer loop are performed already. 
    The first one  with only the external part of the potential perturbation in the Hamiltonian, which can be understood as constructing a "starting perturbation", and the second with the first full effective potential perturbation. 
    This is the first density designated to be mixed. 
    We thereby ensure that all density perturbations coming into the mixing procedure are constructed in the same way with the same kind of potential.

\section{State-Dependent Terms of the Dynamical Matrix} \label{sec:app_c1}
    Due to the complexity of the second derivative, a bunch of state-dependent terms appear in the calculation of the DM. With the introduction of matrix-vector products of the (perturbed) expansion coefficients with matrices akin to the Hamiltonian and overlap, the $C$-coefficients from~\eqref{eqn:ccoeffs} can be rearranged into a somewhat compact form that looks as follows:
    \begin{IEEEeqnarray}{rCl}
        C_{\mathbf{k}\nu}^{(1)\alpha i-}&=&\mathbf{z}_{\mathbf{k}\nu}^{\dagger}\cdot\left(\matr{\tilde{H}}^{(1)}(\mathbf{k})-\epsilon_{\mathbf{k}\nu}\matr{\tilde{S}}^{(1)}(\mathbf{k})\right)\cdot\mathbf{z}_{\mathbf{k}\nu}\,, \IEEEyesnumber\IEEEyessubnumber \\
        C_{\mathbf{k}\nu}^{(2)\beta j+\alpha i-}&=&\mathbf{z}_{\mathbf{k}\nu}^{\dagger}\cdot\left(\matr{\tilde{H}}^{(2)}(\mathbf{k})-\epsilon_{\mathbf{k}\nu}\matr{\tilde{S}}^{(2)}(\mathbf{k})-\epsilon_{\mathbf{k}\nu}^{(1)\beta j}\matr{\tilde{S}}^{(1)}(\mathbf{k})\right)\cdot\mathbf{z}_{\mathbf{k}\nu}\IEEEnonumber\\
        &&{}+2\mathbf{z}_{\mathbf{k}\nu}^{\dagger}\cdot\left(\matr{\tilde{H}}^{(1)\dagger}(\mathbf{k}+\mathbf{q})-\epsilon_{\mathbf{k}\nu}\matr{\tilde{S}}^{(1)\dagger}(\mathbf{k}+\mathbf{q})\right)\cdot\mathbf{z}_{\mathbf{k}+\mathbf{q},\nu}^{(1)\beta+} \IEEEyessubnumber \label{eqn:app_eigterms} \\
        &&{}+\mathbf{z}_{\mathbf{k}\nu}^{\dagger}\cdot\matr{\tilde{V}}^{(2)}(\mathbf{k})\cdot\mathbf{z}_{\mathbf{k}\nu}\,, \IEEEnonumber
    \end{IEEEeqnarray}
    The auxiliary matrices we introduce (with omitted superscripts referring to the perturbations) are modified forms of the unperturbed Hamiltonian and overlap. To first order they are
    \begin{IEEEeqnarray}{rCl}
        \tilde{H}_{\mathbf{G'}\mathbf{G}}^{(1)}(\mathbf{k}+\mathbf{q})&=&\mathrm{i}(G-G'-q)_{i}\Braket{\phi_{\mathbf{k} + \mathbf{G}' + \mathbf{q}}|\mathscr{H}|\phi_{\mathbf{k}+\mathbf{G}}}_{\!\!\alpha}+\Braket{\phi_{\mathbf{k} + \mathbf{G}' + \mathbf{q}}|\Theta_{\mathrm{IR}}^{(1)\alpha i-}\mathscr{T}|\phi_{\mathbf{k}+\mathbf{G}}},\IEEEnonumber \\
        \tilde{S}_{\mathbf{G'}\mathbf{G}}^{(1)}(\mathbf{k}+\mathbf{q})&=&\mathrm{i}(G-G'-q)_{i}\Braket{\phi_{\mathbf{k} + \mathbf{G}' + \mathbf{q}}|\phi_{\mathbf{k}+\mathbf{G}}}_{\!\!\alpha}+\Braket{\phi_{\mathbf{k} + \mathbf{G}' + \mathbf{q}}|\Theta_{\mathrm{IR}}^{(1)\alpha i-}|\phi_{\mathbf{k}+\mathbf{G}}}\IEEEyesnumber \,,
    \end{IEEEeqnarray}
    where the main modification is given by a prefactor stemming from the basis variations. The same prefactor, albeit in the other perturbation direction, again modifies the matrices to second order:
    \begin{IEEEeqnarray}{rCl}
        \tilde{H}_{\mathbf{G'}\mathbf{G}}^{(2)}(\mathbf{k})&=&\mathrm{i}(G-G')_{j}\tilde{H}_{\mathbf{G'}\mathbf{G}}^{(1)}(\mathbf{k})\mathrm{\delta}_{\beta\alpha}\,,\IEEEnonumber\\
        \tilde{S}_{\mathbf{G'}\mathbf{G}}^{(2)}(\mathbf{k})&=&\mathrm{i}(G-G')_{j}\tilde{S}_{\mathbf{G'}\mathbf{G}}^{(1)}(\mathbf{k})\mathrm{\delta}_{\beta\alpha}\,,\IEEEyesnumber \\
        \tilde{V}_{\mathbf{G'}\mathbf{G}}^{(2)}(\mathbf{k})&=&\mathrm{i}(G-G')_{i}\Braket{\phi_{\mathbf{k} + \mathbf{G}' }|V_{\mathrm{eff}}^{(1)\beta j+\nabla}|\phi_{\mathbf{k}+\mathbf{G}}}_{\!\!\alpha}\IEEEnonumber \,.
    \end{IEEEeqnarray}
    
An additional derivation akin to~\ref{sec:app_de} holds true  for the usage of the coefficients in the second line of~\eqref{eqn:app_eigterms}, where the exact same coefficients from the Sternheimer loop can be used, albeit resulting in one last additional term
\begin{IEEEeqnarray}{rCl}
    C_{\mathbf{k}\nu,\mathrm{add}}^{(2)\beta j+\alpha i-}&=&
        2\mathbf{z}_{\mathbf{k}\nu}^{\dagger}\cdot\matr{\tilde{S}}^{(1)\dagger}(\mathbf{k}+\mathbf{q})\cdot\mathbf{z}_{\mathbf{k}+\mathbf{q},\nu,\mathrm{add}}^{(1)\beta+}\,, \IEEEyesnumber\IEEEyessubnumber \label{eqn:app_lastterm} \\
    \mathbf{z}_{\mathbf{k}+\mathbf{q},\nu,\mathrm{add}}^{(1)\beta+}&=&-\sum_{\nu'}\frac{1}{2}F(\epsilon_{\mathbf{k}+\mathbf{q},\nu'})\left(H_{\mathbf{q}\nu',\mathbf{k}\nu}^{(1)\beta j+}-\epsilon_{\mathbf{k}+\mathbf{q},\nu'}S_{\mathbf{q}\nu',\mathbf{k}\nu}^{(1)\beta j+}\right)\,, \IEEEyessubnumber \label{eqn:app_lastz1}
\end{IEEEeqnarray}
where, again, for $\delta_{\mathbf{q}\nu',\mathbf{k}\nu}\approx0$ we find $F(\epsilon_{\mathbf{k}+\mathbf{q},\nu'})\rightarrow 1$ and $\epsilon_{\mathbf{k}+\mathbf{q},\nu'}\rightarrow\epsilon_{\mathbf{k}\nu}$.

\section{Calculating the Perturbed Occupation Numbers} \label{sec:app_occ}
    The perturbed occupation numbers $\tilde{f}_{\mathbf{k}\nu}^{(1)\beta j}$ are analytically derived from their original definition
    \begin{IEEEeqnarray}{rCl}
        \tilde{f}_{\mathbf{k}\nu}=f_{\mathbf{k}}F(x),~~~~x=\frac{\epsilon_{\mathbf{k}\nu}-E_{\mathrm{F}}}{k_{\mathrm{B}}T}\,,\label{eqn:app_f0}
    \end{IEEEeqnarray}
    where the smearing function $F(x)$ is taken as the Fermi-Dirac-distribution 
    \begin{IEEEeqnarray}{rCl}
        F(x)&=&\frac{1}{e^{x}+1}\,,
    \end{IEEEeqnarray} with the smearing temperature $k_{\mathrm{B}}T$ in units of the Boltzmann constant $k_{\mathrm{B}}$.
    By taking the derivative of~\eqref{eqn:app_f0} and doing some arithmetic, one can find
    \begin{IEEEeqnarray}{rCl}
        \tilde{f}_{\mathbf{k}\nu}^{(1)\beta j} = -\tilde{f}_{\mathbf{k}\nu}F(-x)\frac{\epsilon_{\mathbf{k}\nu}^{(1)\beta j}-E_{\mathrm{F}}^{(1)\beta j}}{k_{\mathrm{B}}T}\label{eqn:app_f1} \,.
    \end{IEEEeqnarray}
    For the calculation of $E_{\mathrm{F}}^{(1)\beta j}$, there are two options.
    A straight-forward way is to iteratively determine the Fermi energy derivative in the same vein as the Fermi energy itself resulting from the ground-state calculation.
    We instead aim for another analytical scheme that stems from the requirement of a conserved electron count 
    \begin{IEEEeqnarray}{rCl}
        N = \sum_{\mathbf{k}\nu}\tilde{f}_{\mathbf{k}\nu} \,.
        \label{eqn:app_H_N1}
    \end{IEEEeqnarray}
    Once again differentiating both sides using~\eqref{eqn:app_f1}, and rearranging terms leads to
    \begin{IEEEeqnarray}{rCl}
        E_{\mathrm{F}}^{(1)\beta j}=\frac{\sum_{\mathbf{k}\nu}\tilde{f}_{\mathbf{k}\nu}F(-x)\epsilon_{\mathbf{k}\nu}^{(1)\beta j}}{\sum_{\mathbf{k}\nu}\tilde{f}_{\mathbf{k}\nu}F(-x)} \,.
    \end{IEEEeqnarray}
    It was taken into account that according to \eqref{eqn:rho1BasicTRS}, the variation of the left side of \eqref{eqn:app_H_N1} in terms of an atomic displacement is zero, $N^{(1)}=0$. In the case of low smearing, as for insulators, the Fermi energy derivative is taken to be~0.

\section*{References}
\bibliography{main_reply.bib}
\end{document}